%% file: main.tex
\documentclass[twocolumn, tighten, apj, numberedappendix]{openjournal}
\input{defs}

\begin{document}

\title{SCAT Data Release 1: \numSpectra optical spectra of \numSources transients}

\shorttitle{SCAT DR1}
\shortauthors{The SCAT Team}

\input{AUTHORS2.tex}



\begin{abstract}

We present the first data release (DR1) of the Spectroscopic Classification of Astronomical Transients (SCAT) survey, covering the first $\approx 5$~years of observations (March 2018 -- January 2023). DR1 includes \numSpectra spectra of \numSources transients, which we sort into broad spectroscopic classes including supernovae (SNe), transients originating in galactic nuclei, and stellar variability. We collect multi-filter light curves from imaging surveys and fit them with phenomenological models to estimate peak brightnesses and the time of explosion/first-light. Extragalactic transients are matched to candidate host galaxies, and we compare host-galaxy luminosities and projected offsets by SN type. SNe appear to be a reliable way to augment the redshift coverage of nearby ($z\lesssim 0.1$) galaxies in tandem with dedicated redshift surveys. We present new redshifts for roughly half of the SN host galaxies, most of which are low-luminosity dwarfs similar to the Magellanic Clouds ($M_r \gtrsim -18$~mag). This set of transient spectra, light curves, luminosities, redshifts, and host galaxies offers an excellent testbed for real-time photometric/light curve classification pipelines in the modern era of deep and large-area surveys. We conclude with a brief discussion of the provided data products and status of the SCAT survey.

\end{abstract}

\keywords{Transient sources (1851), Spectroscopy (1558), Supernovae (1668), Close binary stars (254), Galaxies (573), Sky surveys (1464)}


\section{Introduction} \label{sec:intro}

Our collective ability to systematically survey the night sky has progressed significantly over the past 3 decades. Improvements in detector technology and computer processing capabilities enable the collection and processing of petabytes of data in near-real time. This has revolutionized the field of time-domain astronomy, especially the discovery and characterization of `transients' -- objects in the night sky that display significant changes in brightness on (relatively) short timescales. Supernovae (\sne) are the canonical transients, but the term has expanded to encompass extreme stellar phenomena (flares, eruptions, outbursts) and physics unique to galactic nuclei and the supermassive black holes (SMBHs) within them. 

Early imaging surveys had limited capacity to collect, process, and store data and thus typically focused on specific science objectives. Zwicky started the first coordinated surveys of nearby galaxies in search of \sne \citep[e.g., ][]{zwicky1964}. The focus on nearby galaxies was continued by projects like the Calan/Tololo Supernova Search \citep{calan_tololo} and the Lick Observatory Supernova Search \citep[LOSS; ][]{loss}, while projects searching for distant \sneia to constrain cosmological expansion, like the Supernova Cosmology Project \citep{scp1, scp2, scp3} and the High-z Supernova Team \citep{highz_team, highz_2}, avoided nearby galaxies and the Galactic plane to keep search fields relatively clean. Surveys dedicated to finding potentially hazardous Solar system bodies, like Spacewatch \citep{spacewatch1, spacewatch2} and the Near-Earth Asteroid Tracking (NEAT, \citealp{neat_paper}) project, prioritized observing along the ecliptic. The All-Sky Automated Survey \citep[ASAS, ][]{asas1, asas2} was one of the first efforts to survey an entire hemisphere across multiple epochs, mostly in search of variable stars. The success of these focused surveys, coupled with further improvements in computer processing, enabled the next generation of larger and deeper surveys like the Catalina Sky/Real-time Transient Surveys (CSS/CRTS; \citealp{css, crts}) and the (intermediate) Palomar Transient Facility (i/PTF, \citealp{ptf_paper1, ptf_paper2}). 

These early surveys evolved into the modern landscape of untargeted, wide-field imaging surveys available today. Smaller projects like the All-Sky Automated Survey for SuperNovae \citep[ASAS-SN, ][]{asassn1, asassn2} and the Evryscope \citep{evryscope1, evryscope2} prioritize cadence and sky coverage by using small ($\approx 10$~cm) telescopes and a single photometric filter. Larger programs, like the Panoramic Survey Telescope and Rapid Response System (Pan-STARRS; \citealp{panstarrs}) and the Zwicky Transient Facility \citep[ZTF, ][]{ztf_paper1, ztf_paper2}, use meter-class mirrors to push deeper across multiple filters but typically require several nights to revisit the same field. Between these are `intermediate' surveys like the Asteroid Terrestrial-impact Last Alert System (ATLAS; \citealp{atlas1, atlas2}) and the Gravitational-wave Optical Transient Observer \citep[GOTO, ][]{goto1, goto2} that use 0.5-m class telescopes to trade depth for cadence. Even more surveys have commenced in just the past few years (e.g., BlackGEM, \citealp{blackgem1, blackgem2}; WFST, \citealp{wfst}; LAST, \citealp{last1, last2}; LS4, \citealp{ls4}). 

The Vera C. Rubin Observatory's Legacy Survey of Space and Time (LSST, \citealp{lsst_paper}), equipped with an effective aperture of $\approx 6.4~\rm m$, marks the culmination of a paradigm shift in time-domain astronomy unfolding over the past 20 years. LSST is expected to discover $\mathcal{O}(10^6)$ transients over its 10-year survey, more than all current and previous surveys combined. This led to significant investments in accurately simulating the unprecedented LSST discovery stream (e.g., \textsc{plasticc}, \citealp{plasticc_paper1, plasticc_paper2, plasticc_paper3}) and, subsequently, filtering that discovery stream with custom `brokers' like \textsc{antares} \citep{antares_paper1, antares_paper2}, \textsc{alerce} \citep{alerce_paper1, alerce_paper2, alerce_paper3, alerce_paper4}, \textsc{fink} \citep{fink_paper1, fink_paper2, fink_paper3}, and \textsc{lasair} \citep{lasair}. Beyond brokers, there is a growing ecosystem for classifying transients using only photometric data (light curves + imaging) and extracting physical quantities (e.g., \citealp{boone2019, villar2020, stausbaugh2022, superphot_paper}).

Extracting maximal information from photometric data is needed to reduce the burden on spectroscopic resources, which have become the bottleneck in follow-up efforts. Even now, only $\approx 10\%$ of transients are classified spectroscopically by all current classification surveys combined (e.g., ePESSTO+, \citealp{pessto_paper}; ZTF BTS, \citealp{ztfbts_paper1, ztfbts_paper2}; SCAT, \citealp{tucker2022}). Spectroscopic classification efforts reliably classify transients reaching $\lesssim 18$~mag, but spectra are rarely obtained for transients fainter than $\gtrsim 20$~mag without flagging by a photometric screening algorithm (e.g., \citealp{needle_paper, elephant_paper}). But, well-curated spectrophotometric samples of transients are crucial to train the photometric classification and inference pipelines. Moreover, spectra uniquely encode abundance, excitation, and velocity information as functions of time, which are essential ingredients for physical models. 

The SCAT survey was conceived in early 2018 \citep{scat_atel} to help convert the increasing number of transients discovered by the world's surveys into tangible science. SCAT has produced numerous publications over the past 8 years of operations that span the breadth of time-domain astronomy including \sneia \citep{derkacy2023, pearson2024, siebert2024, hoogendam2025a, hoogendam2025b, bose2025}, Type II \sne \citep[SNe~II, ][]{tucker2024, baron2025, medler2025, derkacy2026}, stripped-envelope \sne \citep[SESNe, ][]{fraser2021, moore2023, ertini2023, dong2024, shi2026}, tidal disruption events \citep[TDEs, ][]{holoien2019, hinkle2021, hinkle2023, hinkle2024, hinkle2025, hoogendam2024}, ambiguous nuclear transients \citep[ANTs, ][]{neustadt2020, hinkle2022, pandey2025}, active galactic nuclei \citep[AGNs, ][]{neustadt2023}, extreme stellar phenomena \citep{tucker2018, hodapp2020, hodapp2024, aydi2026}, interstellar objects \citep{seligman2025, hoogendam2025c, medler2026}, and more \citep{nicholl2023, gillanders2025, lemon2026}.

\begin{figure*}
    \centering
    \includegraphics[width=\linewidth]{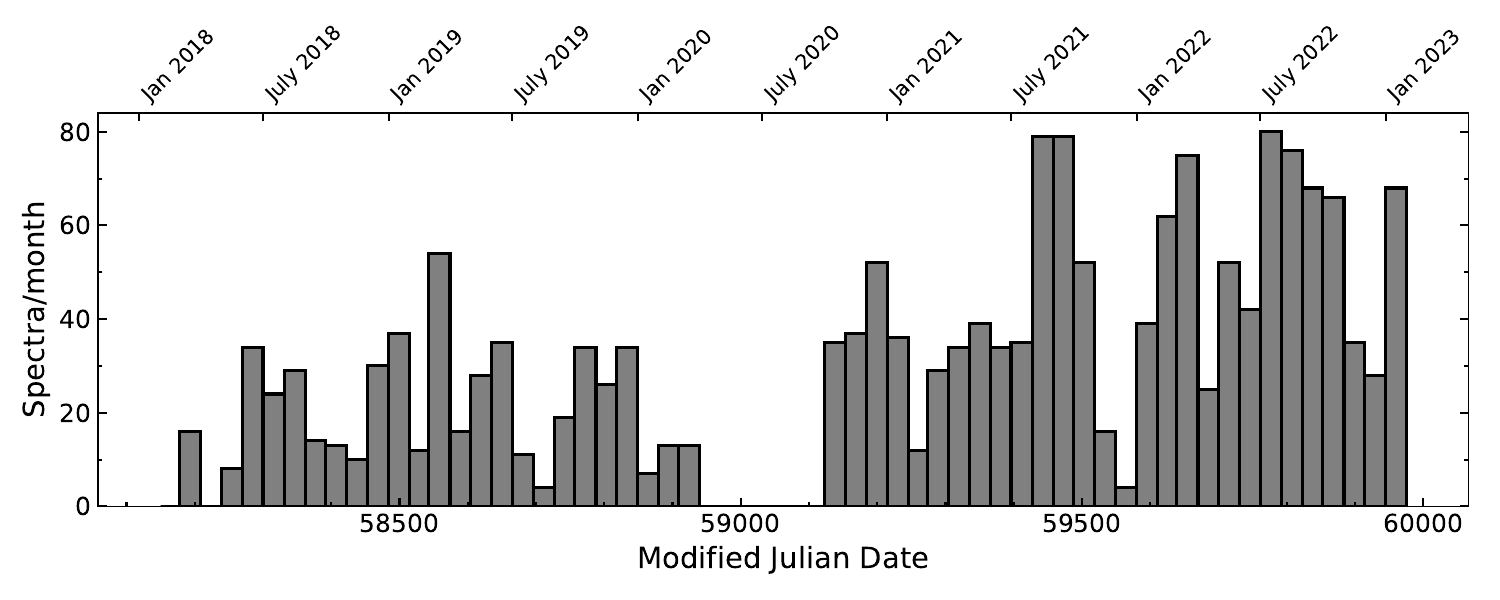}
    \caption{The number of spectra per month for the period covered by DR1. The gap around MJD $\sim 59000$ is due to COVID-19 closures.}
    \label{fig:spec_per_month}
\end{figure*}

In this manuscript, we describe the first data release (hereafter DR1) of the SCAT survey. \S\ref{sec:data} describes the scope of DR1 and updates to the spectroscopic extraction and calibration processes. \S\ref{sec:class} describes the classification approach and taxonomy. \S\ref{sec:hosts} associates the extragalactic transients with potential host galaxies, including redshift and distance estimates. \S\ref{sec:lightcurves} outlines the creation and fitting of the multi-filter light curves from imaging surveys to measure times of first-light and peak brightnesses. \S\ref{sec:summary} summarizes the data products for DR1 and the outlook for future releases. The DR1 data and documentation can be found in the Zenodo repository.\input{footnotes/zenodo} Luminosity measurements correct for foreground Milky Way extinction using the $E(B-V)$ estimates from \citet{schlafly2011}.

\section{SCAT Survey Data}\label{sec:data}

SCAT uses the SuperNova Integral Field Spectrograph (SNIFS, \citealp{Aldering2002, Lantz2004}) mounted on the University of Hawai`i 2.2-m (UH2.2m) telescope to observe transients reported to the Transient Name Server (TNS).\input{footnotes/tns} SNIFS is an integral field unit (IFU) delivering a $6\arcsec\times6\arcsec$ field-of-view (FoV) with $0\farcs4$ spatial pixels (spaxels). The Blue ($\lambda=3200-5200~\AAA$) and Red ($\lambda =5000-9200~\AAA$)\input{footnotes/Rch_H2O}
channels cover the full optical range simultaneously at a resolution of $R = 1200\approx 250$~\kms, although one channel is occasionally offline for maintenance. Details about the survey motivation, operations, and initial data processing are provided in \citet{tucker2022}. Here, we describe the scope of DR1 (\S\ref{subsec:scope}) and updates to our processing and calibration procedures. Improvements to the flat-field alignment and airglow suppression are described in \S\ref{subsec:dichroic} and \S\ref{subsec:airglow}, respectively. We use observations of spectrophotometric standard stars to empirically understand our spectral error limits in \S\ref{subsec:errfloor}. Finally, our qualitative review system is outlined in \S\ref{subsec:scoring}.

\subsection{DR1 Scope}\label{subsec:scope}

\begin{figure*}
    \centering
    \includegraphics[height=2.4in]{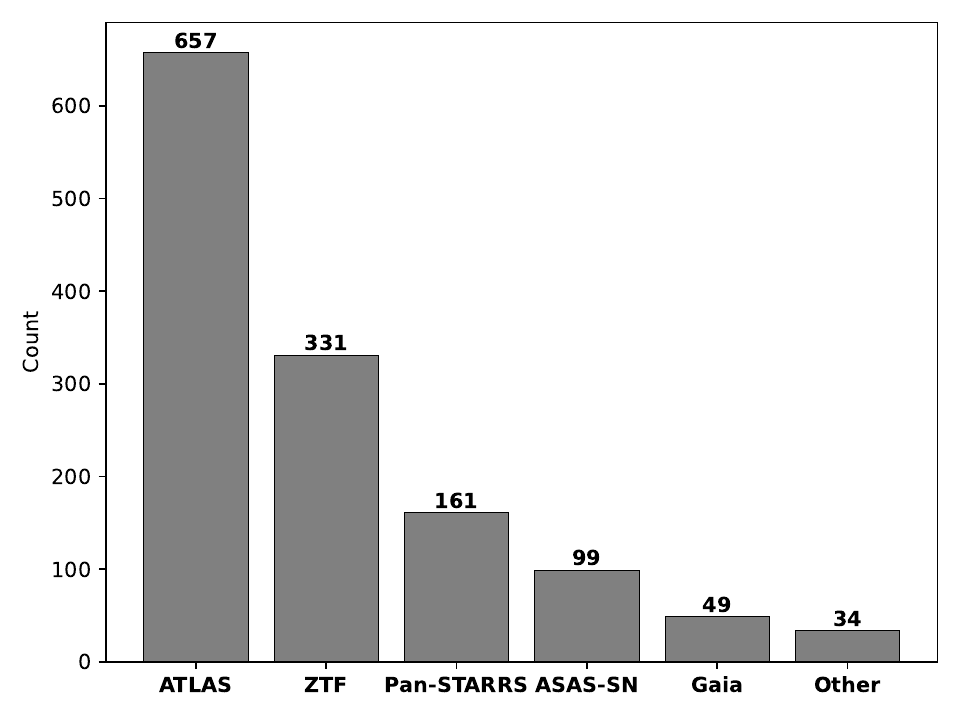}
    \includegraphics[height=2.4in]{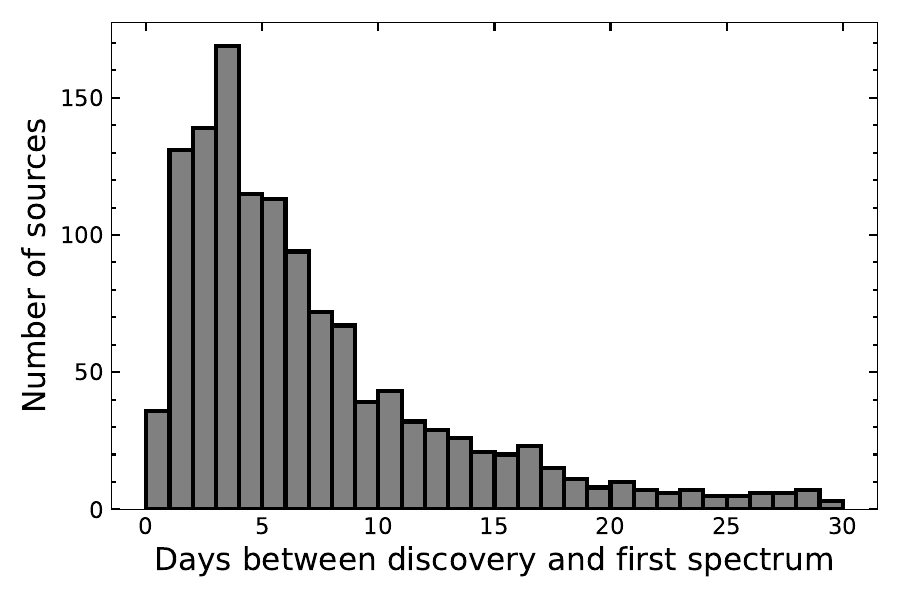}
    \caption{\textit{Left}: Discovery survey distribution. \textit{Right}: Distribution of the time between object discovery (when it was reported to TNS) and our first spectrum.}
    \label{fig:discovery_stats}
\end{figure*}

DR1 includes SNIFS exposures obtained from the start of the survey, March 2018, to the end of the 2022B observing semester (2023--01--31) that targeted a transient reported to TNS. The timeline of science exposures is shown in Fig.~\ref{fig:spec_per_month}. The gradual increase in spectra over time is due to larger telescope allocations and instrumentation/telescope upgrades.

The left panel of Fig.~\ref{fig:discovery_stats} shows the breakdown by discovery survey, including ASAS-SN, ATLAS, ZTF, Pan-STARRS, and \textit{Gaia} Photometric Alerts \citep{gaia_alerts}. The `Other' category includes the Brazilian Transient Search (BraTS) group, the Distance Less Than 40 (DLT40, e.g., \citealp{dlt40_paper}) survey, the Mobile Astronomical System of Telescope Robots (MASTER, \citealp{master_paper}), the Palomar Gattini-IR (PGIR, \citealp{pgir_paper}) survey, the PMO-Tsinghua Supernova Survey (PTSS), the Tsinghua University-Ma Huateng Telescopes for Survey \citep[TMTS, ][]{tmts_paper}, the Xingming Observatory Sky Survey (XOSS), and discoveries by amateur astronomers. 

The right panel of Fig.~\ref{fig:discovery_stats} shows the number of days between an object's discovery (reported to TNS) and our first spectrum. This gives an idea of how quickly we observe most sources after they are discovered by an imaging survey. There is a dependence on the underlying brightness evolution, which is discussed briefly in \S\ref{sec:lightcurves}.

This release encompasses \numSpectra SNIFS spectra of \numSources individual transients, most of which were observed once for spectroscopic classification. DR1 also includes objects flagged by our team for follow-up observations, with 63 sources (4.7\%) having $\geq 3$ epochs and 34 sources (2.6\%) having $\geq 5$ epochs of spectra.\input{footnotes/multispec} We do not combine spectra of the same object taken on the same night to preserve all potential science cases, but combining spectra can trade temporal sampling for improved SNR and artifact rejection. 

This release only includes non-spectrophotometric spectra extracted using point-spread-function (PSF) fitting. Aperture-based extractions might be preferable for some targets with significant host-galaxy contamination, but the vast majority of observations can be adequately extracted with the PSF fitting algorithm as shown in \S\ref{subsec:scoring}. We expect the extraction of highly contaminated sources to improve in the future by leveraging archival multi-filter imaging (see, e.g., \citealp{hypergal} for a similar approach with the SEDMachine), but for now we simply discard observations for which the PSF fitter cannot reliably trace and extract the transient. SCAT already releases aperture-extracted spectra to TNS as part of our public classification program, so DR1 focuses on the spectra with reliable PSF extractions. 

We do not attempt to validate the flux calibration on scales larger than the response spline spacing ($\approx 100-150~\AAA$) because images obtained through the photometric (P) channel are not yet incorporated into our reduction pipelines. The broadband spectral slope is still accurate to $\approx 10\%$ (see \S\ref{subsec:errfloor} and Appendix~\ref{app:errmodel}), although host-galaxy contamination becomes the limiting factor in most science exposures. Absolute spectrophotometry remains a good avenue for future improvements.

\vspace{0.9cm}
\subsection{Mitigating Dichroic Variations}\label{subsec:dichroic}

\begin{figure}
    \centering
    \includegraphics[width=\linewidth]{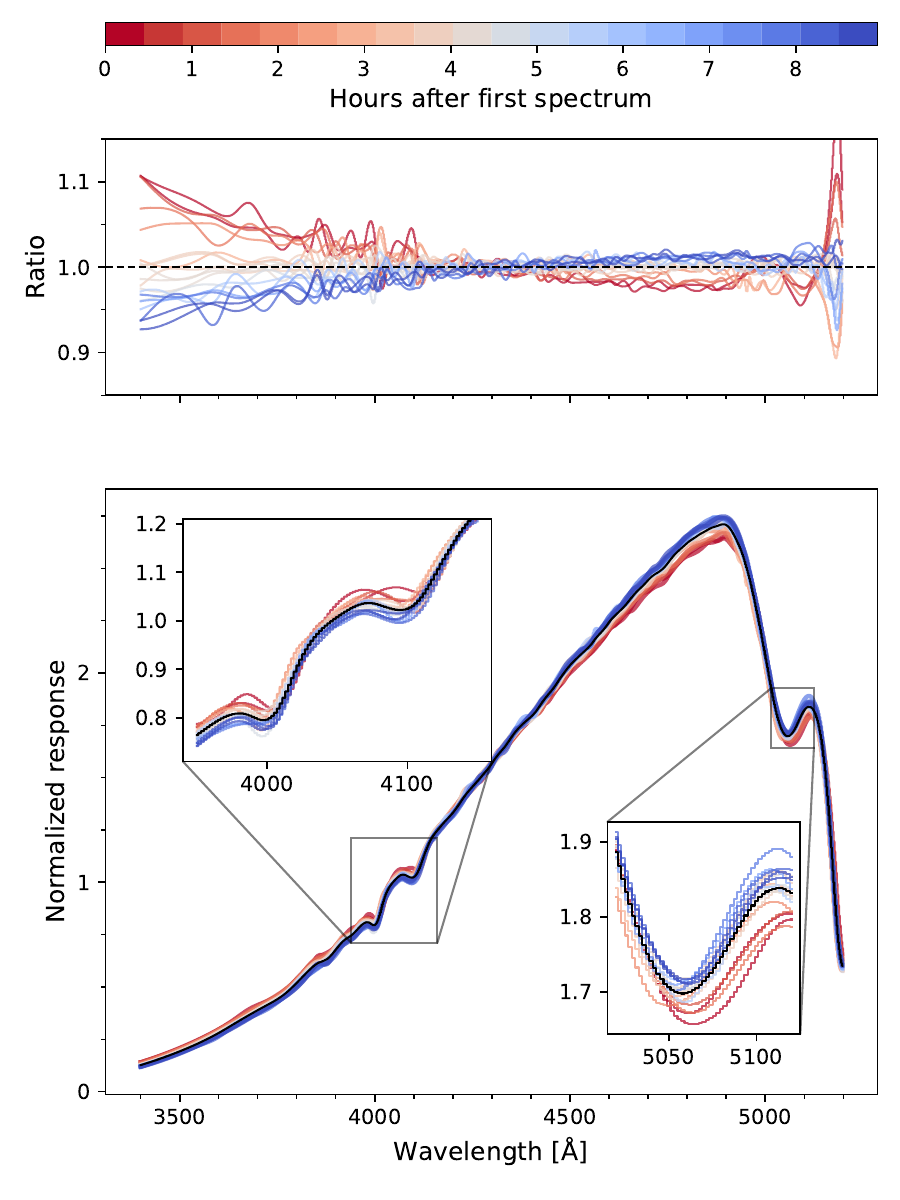}
    \caption{Examples of raster flats from a night with particularly severe variations in dichroic throughput (UT 2021-07-10). This night is at the 95th-percentile of flat-field variability over the course of a night (rms $\approx 1.7\%$). The best nights show variations of rms $\lesssim 0.5\%$. The center-spaxel spectrum for each raster flat is shown in the bottom panel, color-coded by the time elapsed from the first exposure. The mean of the spectra is shown in black. The insets highlight the strong dichroic features near $ 4000~\AAA$ and $ 5000~\AAA$. The upper panel shows the individual flat-field spectra divided by the mean.}
    \label{fig:raster}
\end{figure}

The dichroic beam splitter reflects bluer photons and allows redder photons to pass through, with a gradual transition across $\approx 5000-5200$~\AAA. The SNIFS dichroic has a hygroscopic coating which expands and contracts in response to changes in the ambient humidity, producing $\approx 5-15\%$ throughput variations that shift in wavelength by (up to) $\approx \pm 10~\AAA$. Fig.~\ref{fig:raster} shows the example flat-field spectra for a `bad' night with significant humidity variations. The variable dichroic transmission results in artifacts near $\approx 4000~\AAA$ and complicates the accurate combination of the B and R channels. 

We use a 2--step approach when applying the spectro-spatial flat fields to minimize the effects of this time-dependent throughput. The full-frame spectroscopic flats taken at the beginning of each observing night are combined into nightly `master' flats. Then, we use the raster flat-field frames taken before each science spectrum, which only keep the center SNIFS spaxel, to measure the wavelength offset of the nearest master flat. We found that interpolating the raster wavelength shifts to the midpoint of each exposure produced the best results, as it better captured any shifts during an exposure. 

This process works well for most observations, but nights with significant humidity variations (cf. Fig.~\ref{fig:raster}) are always difficult to correct accurately. Forward-modeling the spectro-spatial flat fields appears promising, but for now, we caution against over-interpreting unexpected P-Cygni-like features near the dichroic regions. These occur when the flat-field is slightly misaligned in wavelength, or there were significant humidity variations that the flat-field templates cannot capture. The misalignment leads to a wavelength-dependent over- or under-correction that manifests as a P-Cygni-like feature. 

We flag and mask any suspicious spectral features during data review. These are also masked in the \textsc{ascii} versions of the spectra, which should be considered the default spectra and are used in all subsequent figures. $\sim 50\%$ of the final spectra have masked-pixel fractions $f_{\rm mask} \geq 5\%$, and $\sim 10\%$ of the spectra have $f_{\rm mask} \geq 10\%$. Our aggressive approach may inadvertently mask valid astrophysical features, so the \textsc{fits}-formatted spectra include the original data. 

\subsection{Empirical Error Model}\label{subsec:errfloor}

\begin{figure}
    \centering
    \includegraphics[width=\linewidth]{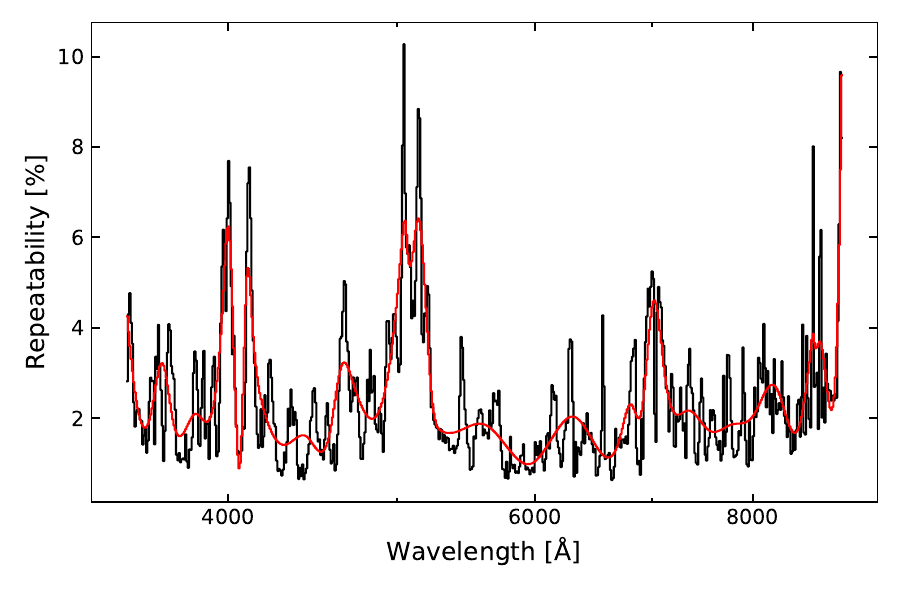}
    \caption{Per-exposure repeatability of our SNIFS spectra from observations of spectrophotometric standard stars (black) with our spline model overlaid (red). The features at $\approx 4100$~\AAA and $\approx 5100$~\AAA are from the dichroic (Fig.~\ref{fig:raster}), and the feature around $\approx 7100$~\AAA is caused by poor continuum modeling during telluric correction. Appendix~\ref{app:errmodel} shows the flux re-normalization for EG~131 as an example.}
    \label{fig:error_floor}
\end{figure}

The DR1 spectra include per-pixel flux uncertainties estimated by the PSF fitting algorithm from the covariance matrix. For bright sources, the reported statistical signal-to-noise ratio (SNR) can easily exceed systematic uncertainties from our data processing and calibration pipelines. Sources of systematic uncertainties include poor background and sky estimation, incorrectly modeled PSF shapes, or poorly-aligned spectroscopic flat-field cubes (cf. \S\ref{subsec:dichroic}). In practice, any effect acting on scales smaller than the instrumental response spline spacing ($100-150~\AAA$) will contribute to the systematic uncertainty floor. 

We create an empirical error model from the white dwarf spectrophotometric standard stars EG~131, G191--B2B, Feige~67, Feige~110, GD~71, and GD~153. These stars were selected for their longer exposure times, mimicking most of our science spectra, smooth blackbody-like spectra, and for having reference spectra in the CALSPEC database \citep{calspec1, calspec2}. Spectra are first rebinned to $10~\AAA$/pixel and then scaled to match the reference CALSPEC spectra. The ratios of the SNIFS and CALSPEC spectra provide a preliminary estimate of per-exposure calibration repeatability of $\approx 10\%$ (see Appendix~\ref{app:errmodel}) with increasing uncertainties at shorter wavelengths from variations in atmospheric throughput and worse flat-fielding in the B channel.

To better capture the true systematic error floor, we correct for any errors in the instrumental and atmospheric response curves using a 7th-order polynomial fit minimizing the median absolute deviation (MAD) to reduce the effects of outliers. This does a decent job of removing large-scale features without compromising the smaller-scale features we are trying to capture. Fig.~\ref{fig:error_floor} shows the derived empirical error floor from this process which should be a good estimate of the calibration repeatability. This empirical error model is added in quadrature to the statistical errors estimated from the PSF extraction. The error model has a mean (median) of 2.4\% (2.0\%).\input{footnotes/snifs} An example derivation of the error model for one standard star (EG131) is included in Appendix~\ref{app:errmodel}. 

\subsection{Airglow Subtraction}\label{subsec:airglow}

The PSF fitting algorithm often struggles to accurately estimate the background when the host is within a few arcseconds of the transient and of comparable brightness. This can leave noticeable [\ion{O}{1}]$\lambda5577,6300~\AAA$ airglow emission in the spectrum due to confusion between the sky and spatially-extended structure from the host galaxy. This source of uncertainty will not be captured by the empirical error model because the calibration stars do not have nearby structure.

We fit residual [\ion{O}{1}] $\lambda5577,6300~\AAA$ emission with simple Gaussian emission-line models plus flat continua to mitigate the worst cases. The central wavelength is allowed to vary by $\pm 1~\AAA$ (0.25 spectral pixels). The line width is considered a free parameter, and any fits with line widths $\geq 1.5\times$ the spectral resolution of SNIFS ($R=1200\approx 250$~\kms) are rejected as spurious. This simple approach fixes the vast majority of cases without introducing new issues. 

\subsection{Qualitative scoring}\label{subsec:scoring}

\begin{figure*}
    \centering
    \includegraphics[width=\linewidth]{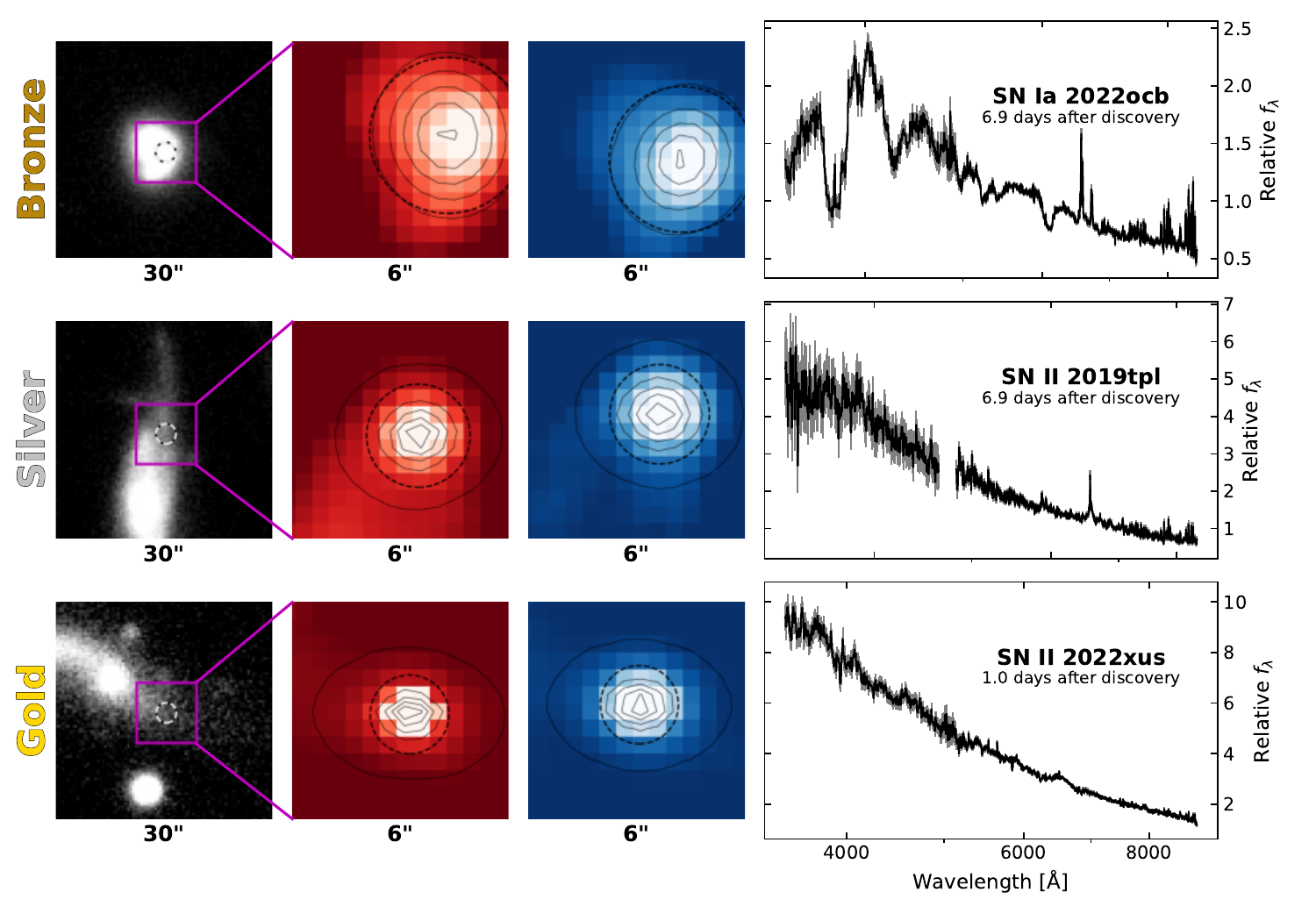}
    \caption{Examples of the different quality scores: Bronze (top row), Silver (middle row), and Gold (bottom row). The left column shows a $30\arcsec\times30\arcsec$ cutout of the local environment with the approximate SNIFS FoV shown in purple. The dashed black and white circle is 2\arcsec in diameter. The middle two rows show the R (middle left; $\lambda=6000-8000~\AAA$) and B (middle right; $\lambda = 3500-4500~\AAA$) images created from the SNIFS IFU cubes with the PSF contours (solid lines) and effective seeing FWHM (dashed circle) overlaid. The positional shifts between the B and R channels are mostly due to atmospheric differential refraction \citep[ADR, ][]{filippenko1990} with a small offset ($\Delta x\approx 0\farcs2$, $\Delta y\approx0\farcs7$) from the inter-channel alignment. The right column shows the final spectrum for each exposure.}
    \label{fig:quality_examples}
\end{figure*}

\begin{figure}
    \centering
    \includegraphics[width=\linewidth]{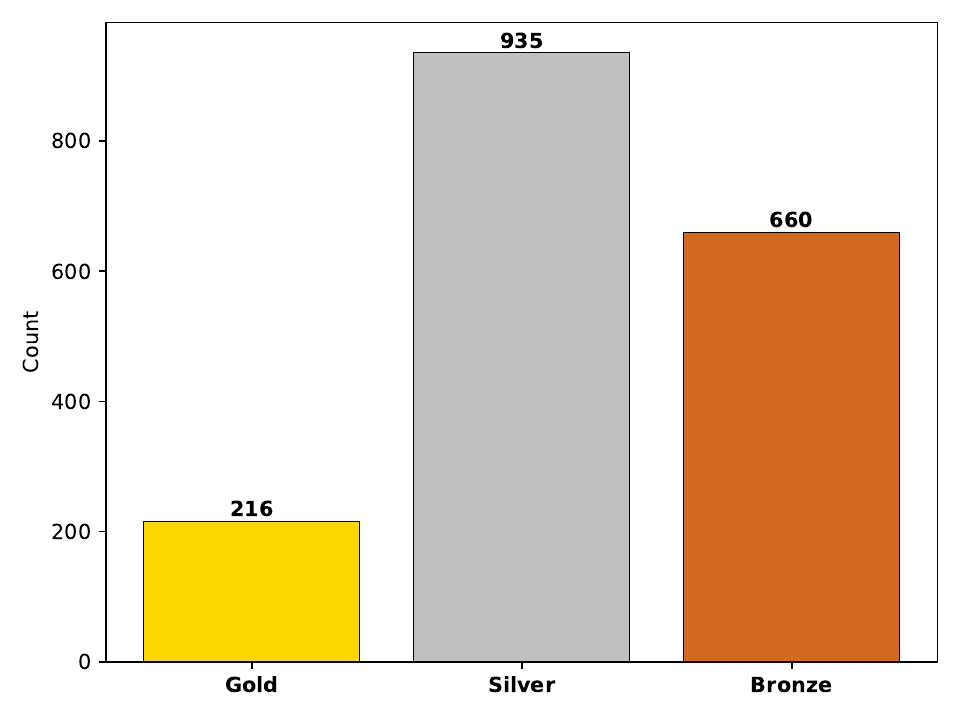}
    \caption{Distribution of spectroscopic quality scores described in \S\ref{subsec:scoring}.}
    \label{fig:quality-pie}
\end{figure}

The final step in preparing the spectra is assigning it one of the \textit{qualitative} scores: Gold, Silver, or Bronze. These are designed to capture the overall extraction quality of each spectrum. Scores were assigned manually after the SCAT data quality review.\input{footnotes/reviewers} Fig.~\ref{fig:quality_examples} provides examples of each class. The PSF extractions of Bronze-class observations often show skewed object traces and/or inaccurate seeing measurements. The breakdown of the different classes is shown in Fig.~\ref{fig:quality-pie}. 

The provided quality scores are intended to guide users on which data products meet their science requirements. Generally, the Bronze exposures have sufficient background contamination that we advise against using these observations for spectral measurements because the continuum is likely affected. There are plenty of Bronze spectra with usable wavelength regions, but users should visually inspect these exposures before analysis to check if the level of contamination meets their science requirements. Silver exposures have minor defects that may introduce small issues with the reductions, most often minor (but visible) host-galaxy contamination and/or bad seeing ($\gtrsim 1\farcs5$), where the PSF becomes difficult to model \citep{rubin2022}. Gold exposures show no obvious defects or issues in the PSF extraction, though we caution that these exposures can still suffer from poor flat-field alignment and dichroic issues. 

The quality scores can correlate with transient properties. For example, stellar sources are more likely to have Gold-quality exposures because they are unlikely to have a nearby galaxy that contaminates the image background. Conversely, extragalactic transients often have a (candidate) host galaxy (see \S\ref{sec:hosts}) which can hinder accurate measurement and subtraction of the sky background. In these cases, the quality score also depends on the (wavelength-dependent) flux ratio of the source to the local surface-brightness of the host, and the ratio of the image quality (atmospheric seeing + PSF) to the projected separation between the source and the host. 

While the provided scores are qualitative by design, the warning and error flags recorded during PSF extraction can be used to quantitatively select spectra. These flags are included in the \textsc{fits} headers of the 1D spectra, and we identify several metrics that correlate with poor exposure quality. Examples include exposures with inconsistent seeing estimates between the B and R channels, inconsistent seeing estimates between the science exposures and the guider video frames, and exposures with discrepant traces (i.e., the object traces do not align across the B and R datacubes). The DR1 documentation provides a full description of the exposure-level flags.

\section{Source Classification}\label{sec:class}

\begin{figure*}
    \centering
    \includegraphics[height=2.4in]{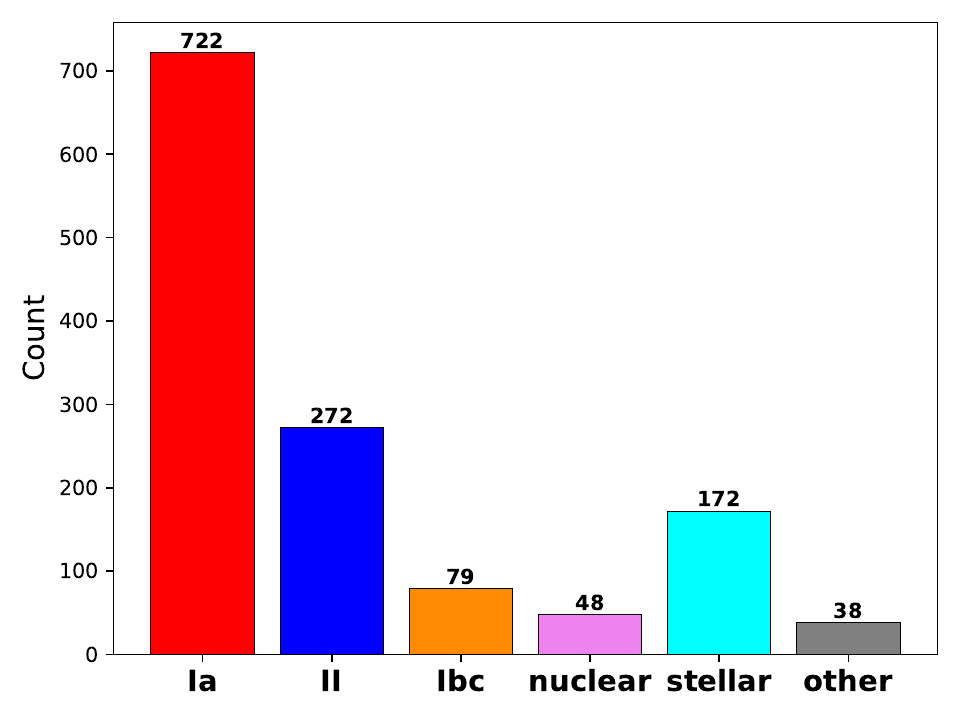}
    \includegraphics[height=2.4in]{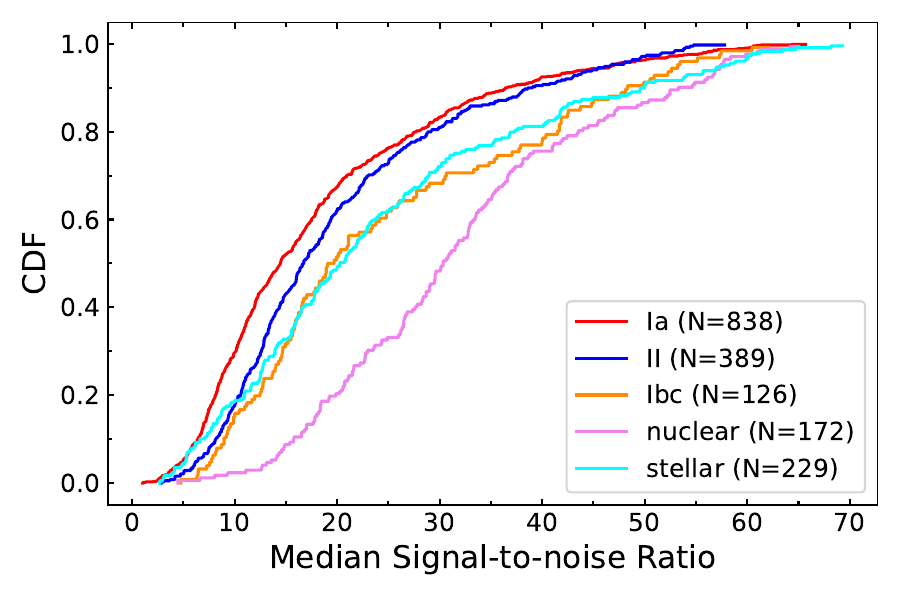}
    \caption{\textit{Left}: The distribution of objects by spectral type in DR1. \textit{Right}: The median spectral S/N, color-coded by spectral type.}
    \label{fig:snr}
\end{figure*}

We attempt to sort each source into 5 empirical classification groups: \sneia, \sneii, \sneibc, nuclear transients, and stellar variability. 
Classifications are done manually using all available data products (spectra, light curves, host associations), with the assistance of template matching for the \sne (\S\ref{subsec:supernovae}). We take prior classifications from literature, where available, and use any publicly-available information (i.e., additional classification spectra on TNS) when the SNIFS spectra are inconclusive. We use the archival imaging to determine the plausibility of potential hosts in \S\ref{sec:hosts}. The survey light curves, described in \S\ref{sec:lightcurves}, are used to determine spectra phases relative to $t_1$, the time of first-light, and \tmax, the time of peak brightness, in addition to measuring the peak luminosity and color evolution.

We use the \textsc{ascii} versions of the SNIFS spectra for classification which include the wavelength masks derived during the data review. We subtract the fitted host-galaxy narrow lines, when present, following the process in \S\ref{subsec:redshifts-distances} to further reduce host-galaxy contamination in the classification results. Final source classifications were reviewed by the team and any uncertain classifications have the `tentative' flag set in the final source table (see Appendix~\ref{app:products}).

The supernovae are collectively analyzed in \S\ref{subsec:supernovae} and split between thermonuclear/Type Ia \sne, H-rich/Type II \ccsne, and H-poor/Type Ibc \ccsne. Transients from galactic nuclei and stellar sources are discussed in \S\ref{subsec:nuclear} and \S\ref{subsec:stellar}, respectively. Any sources that do not fall cleanly into one of these categories are discussed in \S\ref{subsec:other}.

\subsection{Supernovae}\label{subsec:supernovae}
\input{tables/object_counts}
The left panel of Fig.~\ref{fig:snr} shows that \sne comprise the majority ($\sim 80\%$) of the objects in DR1. We split these into the main overarching types: Type Ia/thermonuclear, Type II/H-rich, and Type Ibc/H-poor/stripped-envelope events (see \citealp{sntypes1, sntypes2} for reviews on \sn types). 

\begin{figure}
    \centering
    \includegraphics[width=\linewidth]{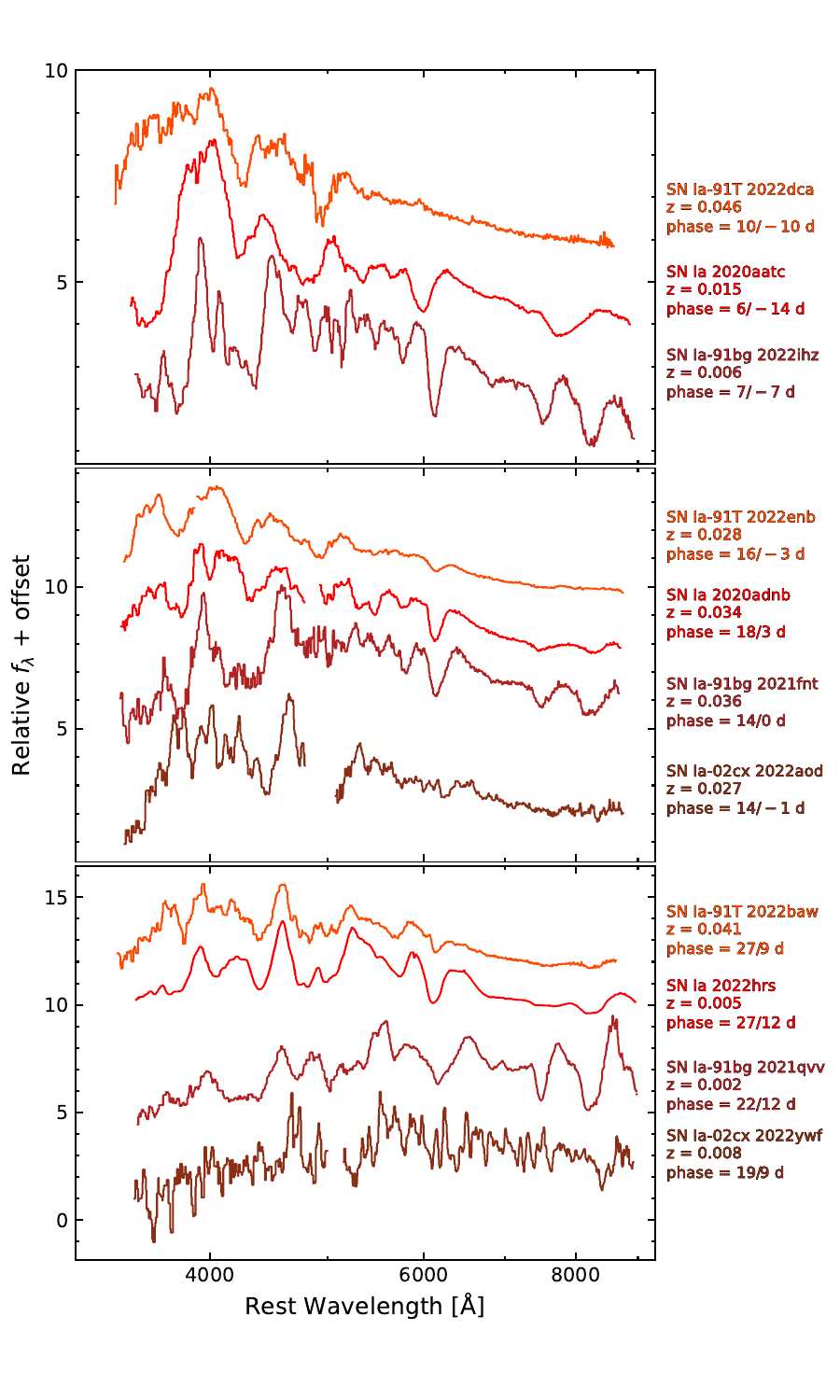}
    \caption{Example spectra of \sneia $1-2$ weeks before peak (top), around peak brightness (middle), and $1-2$ weeks after peak (bottom). Names, redshifts, and phases relative to $t_1$/\tmax are shown on the right side.}
    \label{fig:sneia_examples}
\end{figure}

\begin{figure}
    \centering
    \includegraphics[width=\linewidth]{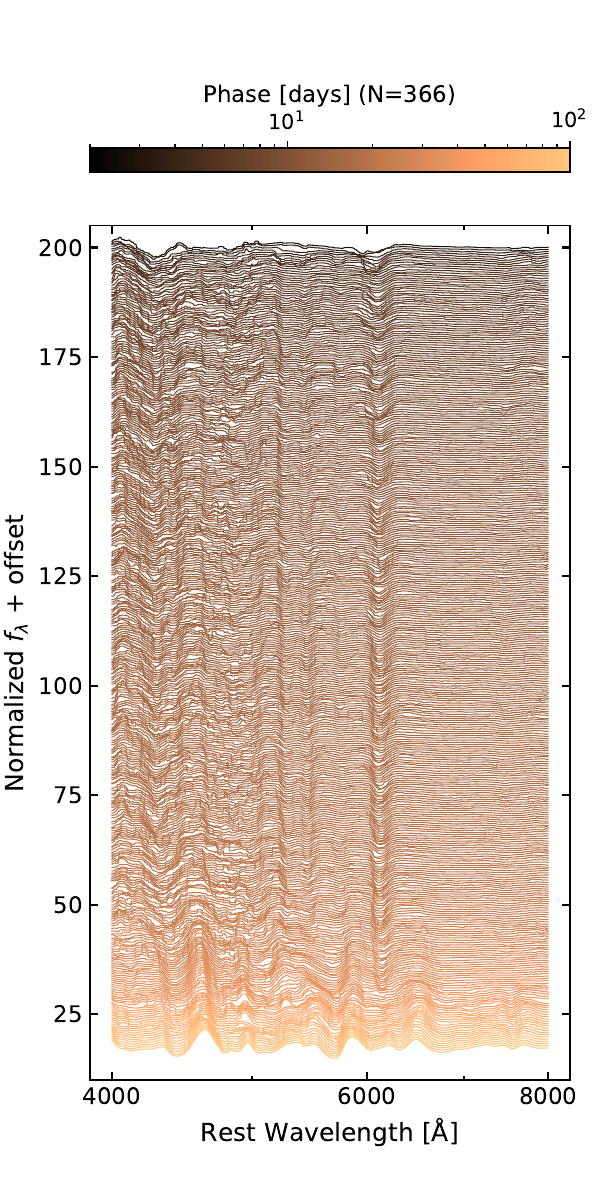}
    \caption{Spectra of `normal' \sneia matching the criteria outlined in \S\ref{subsec:supernovae} and ordered by phase relative to $t_1$.}
    \label{fig:sneia}
\end{figure}

We use the new template-matching software \textsc{snid-sage} \citep[v1.2.2, ][]{sage},\footnote{\url{https://github.com/FiorenSt/SNID-SAGE}} an updated version of the original SuperNova IDentification (SNID) code \citep{blondin2007}, to differentiate the SN (sub-)types. \sage uses the Super-SNID template library \citep{supersnid} plus additional \ccsn templates from the Modjaz METAL group \citep{MetalLib1, MetalLib2, MetalLib3, MetalLib4, MetalLib5, MetalLib6}.

The SNIFS spectra are median-filtered and rebinned to 10~\AAA/pixel before processing with \sage to improve the SNR and minimize artifacts. Masked regions are linearly interpolated over by fitting the 5 spectral pixels on either side of the wavelength mask. If a spectroscopic redshift is known for the host galaxy (see \S\ref{sec:hosts}), the \sage redshift range is limited to $z_{\rm host} \pm0.02$.

We take a conservative approach when assigning a spectroscopic subtype (i.e., Ia-norm vs. Ia-91bg) because many factors can influence accurate subtyping, particularly the phase of the spectrum and its quality. Consequently, we do not use the `norm' suffix to denote normal events, as we may have simply missed the right phase or cannot resolve the required spectral features. Instead, the subtype is simply the spectral type (`Ia-Ia') if we cannot say with confidence that it is a specific subtype.

\begin{figure}
    \centering
    \includegraphics[width=\linewidth]{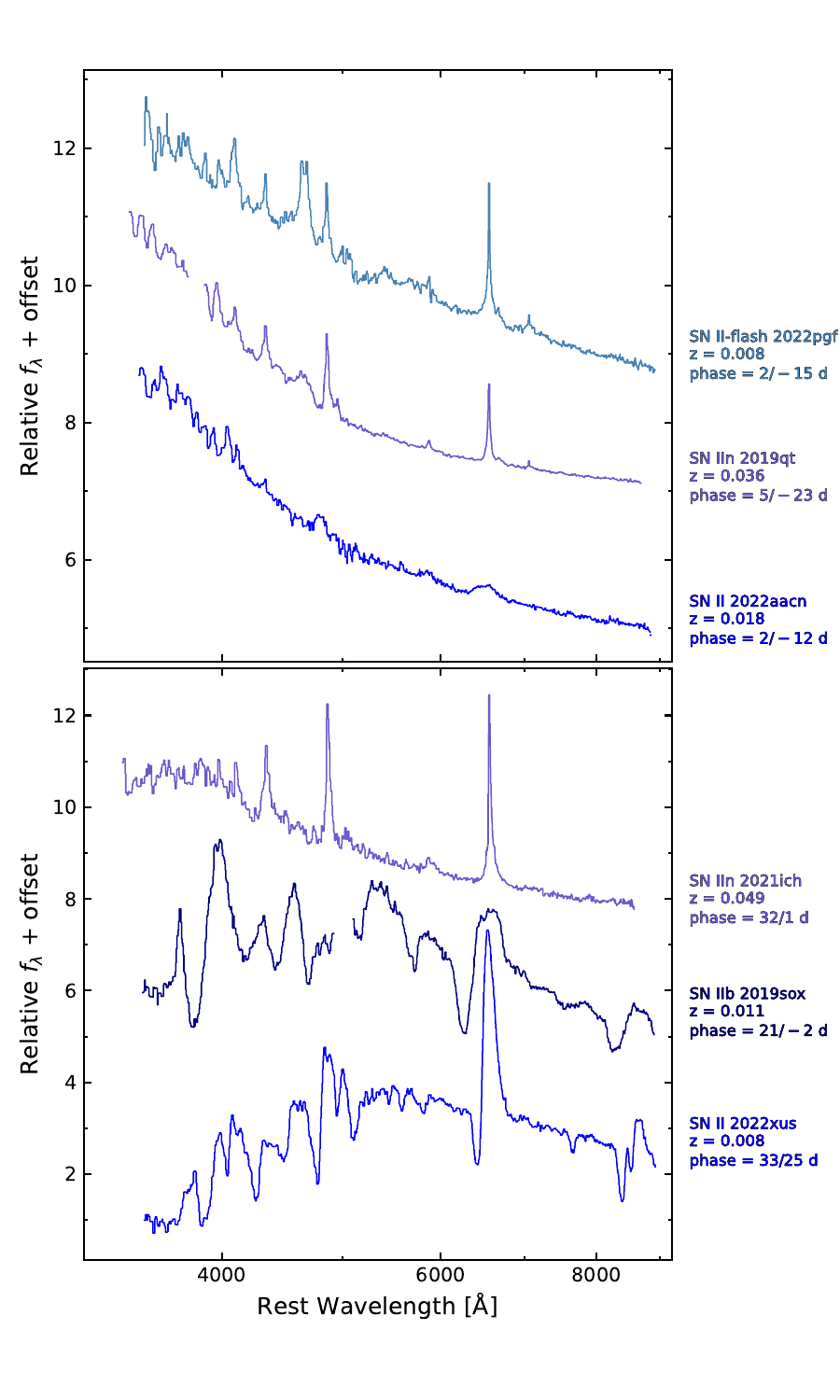}
    \caption{Example spectra of \sneii at early (top) and intermediate (bottom) phases. The name, redshift, and phase relative to $t_1$/\tmax for each spectrum are listed on the right. An earlier spectrum of SN~2022xus is shown in Fig.~\ref{fig:quality_examples}.}
    \label{fig:sneii-examples}
\end{figure}

\begin{figure}
    \centering
    \includegraphics[width=\linewidth]{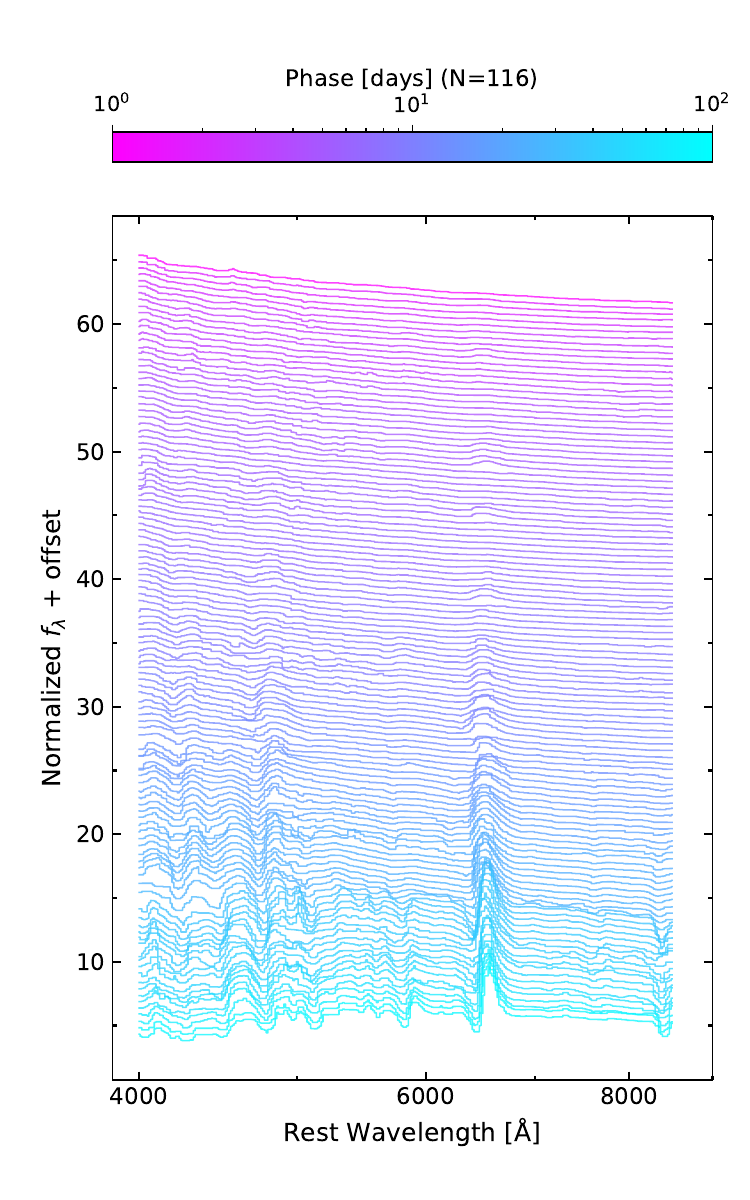}
    \caption{Spectroscopic time-series of \sneii spectra that satisfy the quality cuts in \S\ref{subsec:supernovae}, ordered by phase from top to bottom.}
    \label{fig:sneii}
\end{figure}

\begin{figure}
    \centering
    \includegraphics[width=\linewidth]{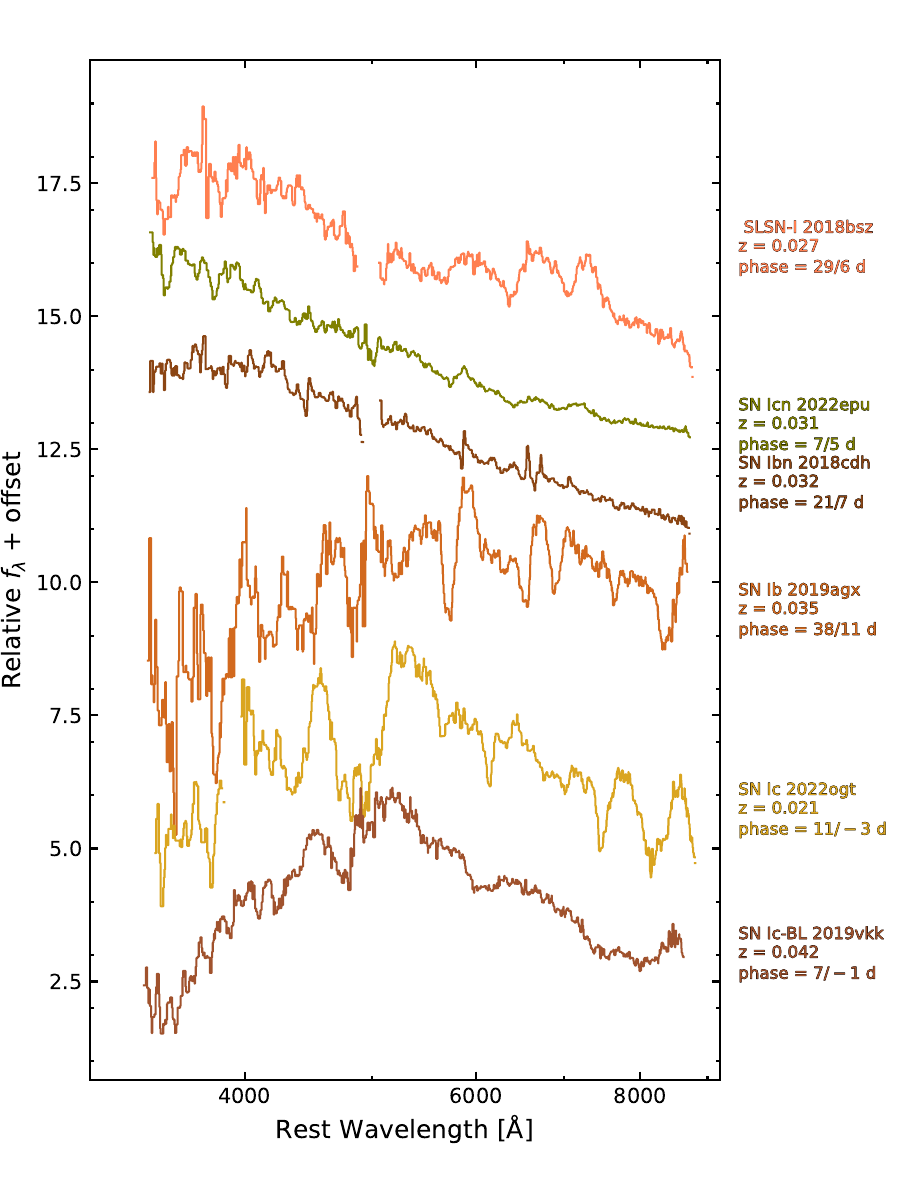}
    \caption{Example spectra of \sneibc around maximum brightness. The names, redshifts and phases relative to $t_1/\tmax$ are given along the right side.}
    \label{fig:sneibc}
\end{figure}

We include 5 \snia subtypes with example spectra shown in Fig.~\ref{fig:sneia_examples}. These include the overluminous 91T-like \sneia \citep{91T_paper1, 91T_paper2} and the underluminous 91bg-like \sneia \citep{91bg_paper1, 91bg_paper2}, which are often considered extensions of the spectroscopically-normal population \citep[e.g., ][]{li2022, obrien2024}. More extreme \snia variants include the very luminous 03fg-like \sneia \citep{03fg-1, 03fg-2}, sometimes referred to as Super-$M_{\rm Ch}$ \sneia due to the above-average luminosities, and the very low-luminosity class of 02cx-like \sneia, often referred to as SNe~Iax \citep{Iax-1, Iax-2}. We have a few 02es-like \sneia \citep{02es} and two cases of a \snia interacting with large amounts of H-rich circumstellar material, referred to as SNe~Ia-CSM or 02ic-like \citep{Ia-CSM1, Ia-CSM2}. Finally, we include the `lensed' category for \snia~2022qmx \citep{22qmx-1, 22qmx-2} because it is $\approx 4$~mag more luminous than the population despite being a relatively normal \snia \citep{22qmx-3, 22qmx-4}. Fig.~\ref{fig:sneia} shows the spectroscopic time-series of normal \sneia with the following quality cuts:

\small
\begin{itemize}
    \item quality $\neq$ Bronze;
    \item median SNR $\geq 5$;
    \item full spectral coverage (B+R channels);
    \item phase $\leq 100$~days after $t_1$;
    \item $t_1$ uncertainty $\leq 5$~days; and
    \item redshift $\leq 0.1$, to keep the rest-frame wavelength ranges consistent. 
\end{itemize}
\normalsize

Next are the \sneii, which represent the explosive death of red supergiants retaining some or all of their H envelopes \citep{smartt2009}. We consider 3 \snii subtypes: SNe~IIb with thin H envelopes \citep{93J1, 93J2}, SNe~IIn interacting with large quantities of H-rich CSM \citep{IIn-1,IIn-2}, and those showing `flash' ionization features around $\approx 4650~\AAA$ from \ion{He}{2}, \ion{C}{3}, and \ion{N}{3} \citep{flash_paper1, flash_paper2}. Examples of these are shown in Fig.~\ref{fig:sneii-examples}.

Fig.~\ref{fig:sneii} shows the full time-series of `normal' \sneii using the same cuts as for \sneia in Fig.~\ref{fig:sneia}. The early spectra of \sneii are essentially pure blackbodies with $T_{\rm eff} \gtrsim 10^4$~K. Balmer P-Cygni profiles and metal absorption features appear later as the \sn cools and transitions to the plateau phase \citep[e.g., ][]{gutierrez2014, gutierrez2017}. We classify transients exploding near a plausible host galaxy, that have early spectra characterized by featureless blue continua, and light curves similar to canonical \sneii (a shock cooling peak followed by a slow decline or plateau, \citealp{anderson2014, valenti2016}), as `tentative' \sneii.

Finally, there are the \sneibc shown in Fig.~\ref{fig:sneibc} which originate from the exposed cores of massive stars \citep{eldridge2013}. These include the H-deficient \sneib and the H- and He-deficient \sneic, which are often collectively referred to as `stripped-envelope' (SE) \sne. SNe Ib and Ic can be difficult to distinguish, especially if the SN is observed several days after peak brightness. There are 2 sources in DR1 that match SN Ib and SN Ic templates reasonably well, and we assign them the generic ‘Ibc’ classification. There are four additional \sneibc subtypes: SNe~Ibn showing narrow lines of He \citep{Ibn-1, Ibn-2, Ibn-3}, SNe~Icn showing narrow carbon emission lines \citep{Icn-1}, the high-velocity broad-line SNe~Ic(-BL) sometimes associated with $\gamma$-ray bursts \citep[e.g., ][]{Ic-GRB-1, Ic-GRB-2}, and Type I superluminous \sne (SLSNe-I, \citealp{slsnei-1, slsnei-2}). \slsnei are sometimes considered distinct from \sneibc, but they likely originate from the explosions of massive stars \citep[e.g., ][]{bucciantini2009, kasen2010} and there may be a continuum between `standard' \sneibc and their superluminous counterparts \citep[e.g., ][]{gomez2022}.

One final point on supernova classification: observations with only B-channel coverage (i.e., the R channel was offline for some reason) only cover $\Delta \lambda = 1700~\AAA$ and do not meet the $2000~\AAA$ coverage requirement of \textsc{snid-sage}. To deal with this issue, we simply pad the red end of the spectrum ($\approx5100-5400~\AAA$) with simulated Gaussian `noise' equal to the mean and standard deviation of the input spectrum. This appears to work reasonably well for the 41 affected spectra, most of which are \sneia.

\subsection{Nuclear Transients}\label{subsec:nuclear}

\begin{figure}
    \centering
    \includegraphics[width=\linewidth]{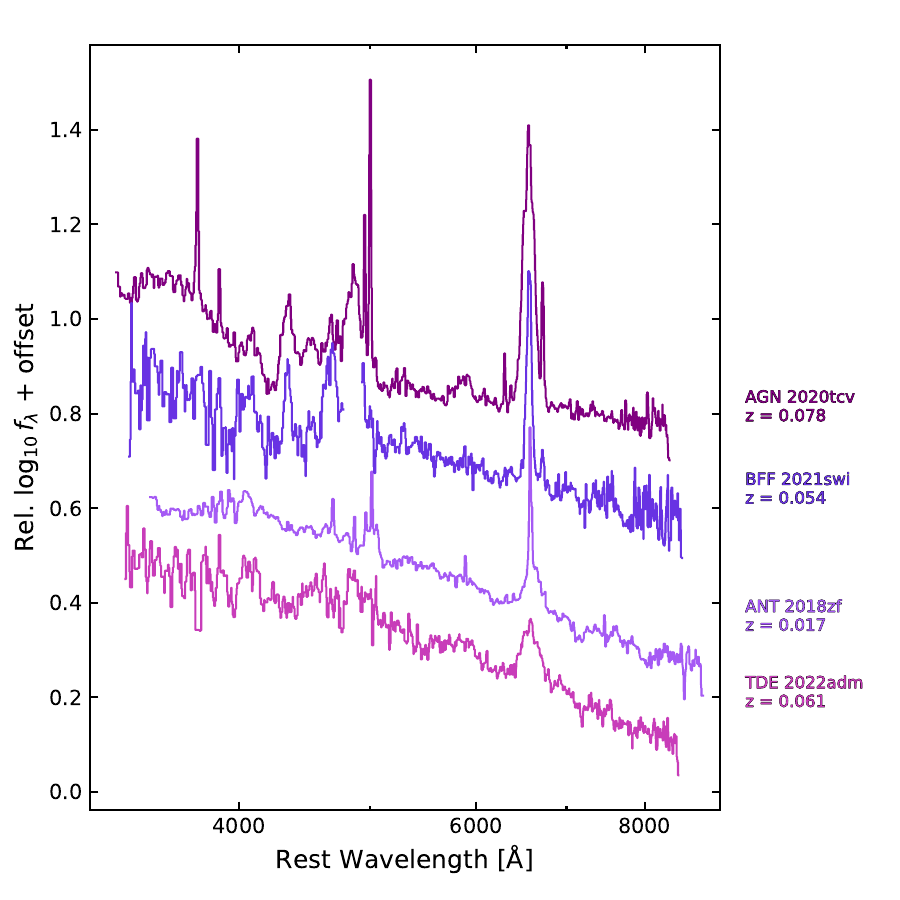}
    \caption{Example spectra for the different types of nuclear transients discussed in \S\ref{subsec:nuclear}. Names and redshifts are provided along the right side.}
    \label{fig:nuclear}
\end{figure}

Changes in the accretion rate onto SMBHs can power a wide variety of time domain phenomena. Given the position of these transients, typically in the center of their host galaxies, we collectively refer to these as nuclear transients and show a subset in Fig.~\ref{fig:nuclear}. AGN show stochastic variability but the AGN typically flagged as transient by imaging surveys are often experiencing larger, more coherent changes in accretion behavior (e.g., `changing-look' AGN, \citealp{clagn}). The emerging class of `Bowen fluorescence flares' \citep[BFF, ][]{bff, bff2} are also thought to represent coherent accretion-rate changes in existing AGNs. 

Distinct from variability caused by changes in an AGN accretion disk, otherwise quiescent SMBHs can also be temporally illuminated by the tidal disruption and subsequent accretion of stars passing within the tidal radius. These are tidal disruption event (TDEs, \citealp{rees1988, evans1989}), where a star is tidally disrupted after passing so close to an SMBH that the tidal forces overwhelm the star's self-gravity.

Observationally, variable AGN spectra generally show broad Balmer emission lines (i.e., Seyfert 1) and narrow forbidden-line emission from ionized gas (e.g., [\ion{O}{3}], [\ion{N}{2}]). BFFs have similar spectra to AGNs, except for unusually strong Bowen fluorescence lines (\ion{He}{2}, \ion{N}{3}, \ion{C}{3}) near $4650~\AAA$. TDEs show persistent blue continua ($T_{\rm eff}\gtrsim 10^4$~K), often with broad emission lines of H and/or He superimposed.
There is also the growing class of `ambiguous nuclear transients' (ANTs, e.g., \citealp{wiseman2025, pandey2026}) showing characteristics of both AGNs and TDEs, highlighting the difficulty of disentangling the source of accretion variations in SMBHs. 

\subsection{Stellar Transients}\label{subsec:stellar}

\begin{figure}
    \centering
    \includegraphics[width=\linewidth]{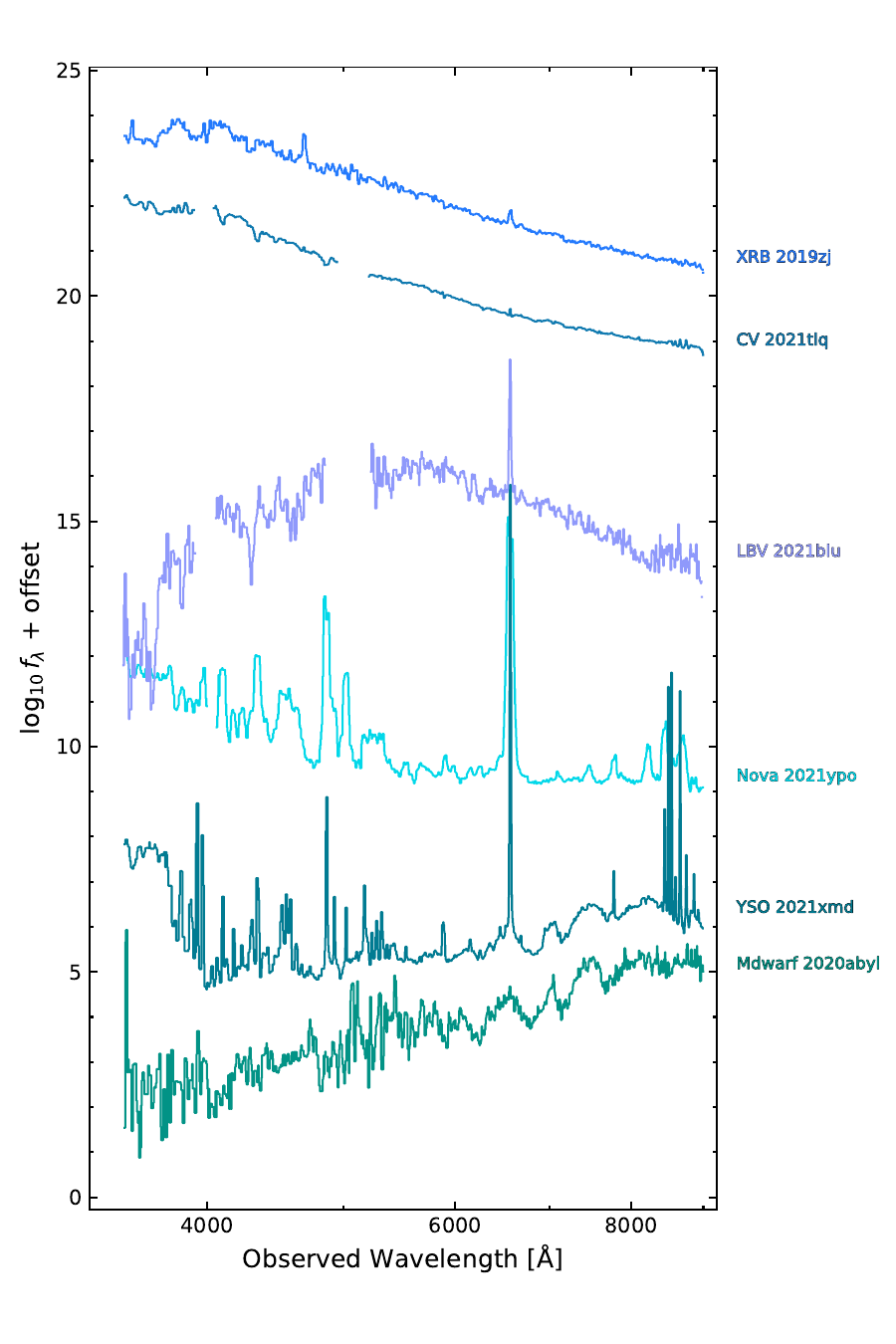}
    \caption{Example spectra of stellar transients. Names are given along the right side.}
    \label{fig:stellar}
\end{figure}

The stellar transients are typically persistent sources experiencing outbursts or flares. Some of these are specifically targeted by SCAT for their uniqueness or rarity, yet most are simply observed during routine classification because we prioritize bright, recently-discovered transients. Even with archival imaging, it can be difficult to determine if a barely-detected smudge is a faint point source or a distant, possibly resolved galaxy. Moreover, several of our stellar sources have galaxies close enough to be considered plausible hosts, and the fast rises can resemble young \ccsne. The number of stellar sources classified by SCAT has declined as we improve our selection function, but some persistent stellar transients are expected given our pursuit of fast-turnaround spectra without waiting for multiple photometric detections. Some of the stellar subtypes are shown in Fig.~\ref{fig:stellar}. We query \textit{Gaia} DR3 \citep{gaiadr3} for all stellar transients and, if a match is found, we use distances from \citet{bailerjones} in our catalog. 

The most numerous of the stellar transients are cataclysmic variable (CV) outbursts triggered by accretion-disk instabilities around a white dwarf \citep{scaringi2026}. Any sources exhibiting blue continua and Balmer lines (emission or absorption) at $z\approx 0$ are considered a likely CV. These are distinct from novae, which are triggered when the accretion layer ignites explosively, and whose spectra show strong emission lines of H, He, and metals \citep{chomiuk2021}. A few of the DR1 novae are in the Milky Way, but the majority are hosted by Andromeda (M31; $z=-0.000991\pm0.000003$, \citealp{1999PASP..111..438F}; $D=770\pm50$~kpc, \citealp{2024AJ....167...31O}).

X-ray binaries (XRBs) have hot blackbody-like spectra like CVs but are distinguished by detectable \HeII, \ion{C}{3}, and/or \ion{N}{3} emission around $\lambda\approx 4650~\AAA$. These high-ionization lines could originate from X-ray irradiation of the donor (Bowen fluorescence, \citealp{bowen}) or the He-rich accretion disk of an AM CVn binary, the H-deficient counterparts to CVs \citep{solheim2010}. At least one XRB in DR1 was detected in outburst by X-ray satellites, the low-mass black hole binary XRB~2018amn (=ASASSN-18ey=MAXI~J1820+070=Gaia18as; \citealp{tucker2018, torres2020}), but we did not systematically check for X-ray counterparts.

Transients with red spectra include flaring low-mass stars, which we labeled as `M~dwarf',\input{footnotes/mdwarf} and outbursts from young stellar objects (YSOs). The strong emission lines of YSOs typically distinguish them from flaring low-mass stars \citep[e.g., ][]{herbig2008}, but the distinction is somewhat qualitative because most YSOs are nascent K/M-dwarfs \citep[e.g., ][]{feigelson1999}. We add the `dipper' classification for AT~2021dzj (Gaia~21bcv), which was flagged as a transient by \textit{Gaia} Transient Alerts due to deep dips in brightness \citep{hodapp2024}.

The final classes of stellar transients in DR1 are Luminous Blue Variables (LBVs) and Luminous Red Novae (LRNe). LBVs are thought to be non-terminal outbursts or eruptions from evolved massive stars \citep{humphreys1994, smith2010}, whereas LRNe are thought to be stellar mergers \citep[e.g., ][]{blagorodnova2017}. These two classes are difficult to distinguish from Intermediate Luminosity Red Transients (ILRTs, e.g., \citealp{cai2021}), which we place under `Other' (\S\ref{subsec:other}) and may represent the very faint end of \ccsn explosions \citep[e.g., ][]{thompson2009, pumo2009}. These three observational classifications show similar spectra, exhibiting weak, moderate-width Balmer emission atop a relatively cool continuum, despite disparate progenitors/origins. We report a few LBVs and LRNe in DR1, but we suggest caution when interpreting these designations.

\subsection{Other/Unknown Sources}\label{subsec:other}

Finally, there are sources that do not fall cleanly into one of the above categories. Roughly half of these are simply low SNR spectra which we cannot reliably match to an existing spectral class, and we label as `Other/unknown'. The other half of them are rare transients and likely \ccsne.

Some events in DR1 are known cases of enigmatic transients defying the taxonomy described above. One example is SN~2018cow, an early, well-studied member of the emerging class of fast blue optical transients (FBOTs, \citealp{cow1, cow2, cow3}).\input{footnotes/cow} Another subset is the `BL Lac' class, which are accreting SMBH systems viewed along the jet axis \citep[e.g., ][]{bllac}. While these could be placed with the nuclear transients in \S\ref{subsec:nuclear}, we place them here because the variability mechanisms are not directly coupled to accretion onto the SMBH and the spectra are distinctly different from all other nuclear transients. As discussed at the end of \S\ref{subsec:stellar}, we also include one ILRT (2022uqn, \citealp{2022uqn}). 

The last subset of sources have early spectra characterized by a hot blackbody with minimal features. These sources are typically projected on-sky close to a plausible host galaxy and do not show the rapid photometric evolution of CVs and XRBs. The early blackbody-like spectra are typically associated with the shock-cooling phase of \ccsne (cf. Fig.~\ref{fig:sneii}) but they are labeled as `other/SN' if (1) there is no redshift, precluding a luminosity estimate, and/or (2) their light curve evolution would be atypical for a \snii. SCAT now attempts to obtain a second classification spectrum of sources exhibiting featureless blue continua to reduce the number of redshift-less, untyped SNe in future releases.

\vspace{0.5cm}
\section{Host Galaxy Associations, Redshifts, Distances, and Luminosities}\label{sec:hosts}

For the extragalactic transients in our catalog, we attempt to (1) associate the source with a candidate host galaxy, (2) measure a redshift, and (3) estimate a distance. 

We take a two-step approach when assigning host galaxies for each transient. First, we use the probabilistic code \prost \citep{prost}\footnote{\url{https://github.com/alexandergagliano/Prost}} to search for potential hosts in the \textsc{Glade+} galaxy catalog \citep{glade1, glade2} and photometric catalogs from the Legacy Surveys \citep[DeCaLS, ][]{dey2019}, and Pan-STARRS DR2 \citep{flewelling2020}. \prost ranks candidate hosts by the transient's offset from the galaxy, normalized by its directional light radius (DLR, e.g., \citealp{gupta2016}). We set an conservative upper bound of $< 10$~DLR on the separation between the transient and potential hosts, and include loose priors on the host-galaxy optical luminosities ($-30 < M_{\rm opt} [\rm{mag}] < -10$) when a redshift is available.

\prost works quite well for $z\gtrsim 0.02$ galaxies but struggles for nearby hosts with large angular sizes that are often `shredded' into multiple small sources in photometric catalogs. Therefore, we also query the NASA Extragalactic Distance Database (NED, \citealp{ned}) for potential host galaxies and visually inspect the NED and \prost matches overlaid on the archival imaging from Pan-STARRS, DeCaLS, or the Digital Sky Survey \citep[DSS, ][]{dss}. These two approaches are highly complementary -- \prost reliably captures the positions and shapes of galaxies smaller than a few arc-minutes but struggles with large, nearby galaxies where NED is very complete. We identify a (candidate) host for $1034/1072$ (96.5\%) of SNe in DR1. 

\subsection{Tentative Hosts}

There are a few cases where the host galaxy association is uncertain. We label the association as `tentative' if it satisfies one of:
\begin{enumerate}
    \item the candidate host is marginally detected (${m_r \gtrsim 23~\rm{mag}}$) by a single survey (Pan-STARRS or DeCaLS), 
    \item the source is projected onto overlapping galaxies (merging/interacting or chance alignment), 
    \item the source is roughly equidistant from two plausible hosts,
    \item the source is offset from a visible host and the available imaging cannot exclude a coincident dwarf galaxy ($M_r \lesssim -18$~mag), or
    \item the source is $\geq 20$~kpc from its assumed host and does not appear to coincide with emission from the host-galaxy. 
\end{enumerate}

\noindent Criteria \#1--3 are relatively straightforward to assess with available archival imaging. Point \#4 occurs when (1) only shallow DSS imaging is available, and/or (2) a low-luminosity dwarf host could remain undetected for higher-redshift sources ($z\gtrsim 0.1$), typically \sneia or SLSNe.

The 91bg-like SN~Ia~2022ihz exemplifies point \#5. The redshift inferred by \sage, averaged over the 8 available spectra, is ${z = 0.0066 \pm 0.0005}$. This agrees with the redshift of the only nearby galaxy, NGC~2974 ($z = 0.006294$), to $0.6\sigma$. Yet SN~2022ihz is offset from the center of NGC~2974 by $\approx 6\farcm7\approx 43$~kpc, or $\sim 17$ effective radii \citep{weijmans2008}. This region of the sky is covered by Pan-STARRS so a coincident dwarf host as faint as $M_r \lesssim -10$~mag is excluded, assuming a conservative detection limit of $r_{\rm ps} = 22$~mag. Thus, we conclude that SN~2022ihz occurred in the very distant outskirts of NGC~2974, representing just a $\sim$few thousandths of the total stellar mass.

This highlights the difficulty of correctly assigning transients to their host galaxies, even when the redshift is known and available imaging is quite sensitive. Some sources are more susceptible to incorrect host-galaxy identification than others. Roughly $\sim 1/3$ of \sneia explode in passive/elliptical hosts \citep[e.g., ][]{rigault2020} where the stellar-mass distribution can extend well beyond the bulk of the optical emission. Conversely, \sneic-BL and SLSNe preferentially explode in metal-poor dwarfs \citep[e.g., ][]{taggart2021} that are more likely to go undetected in survey imaging. In total, we flag 83 tentative SN hosts (8.0\%).

\subsection{Redshifts and Distances}\label{subsec:redshifts-distances}

\begin{figure}
    \centering
    \includegraphics[width=\linewidth]{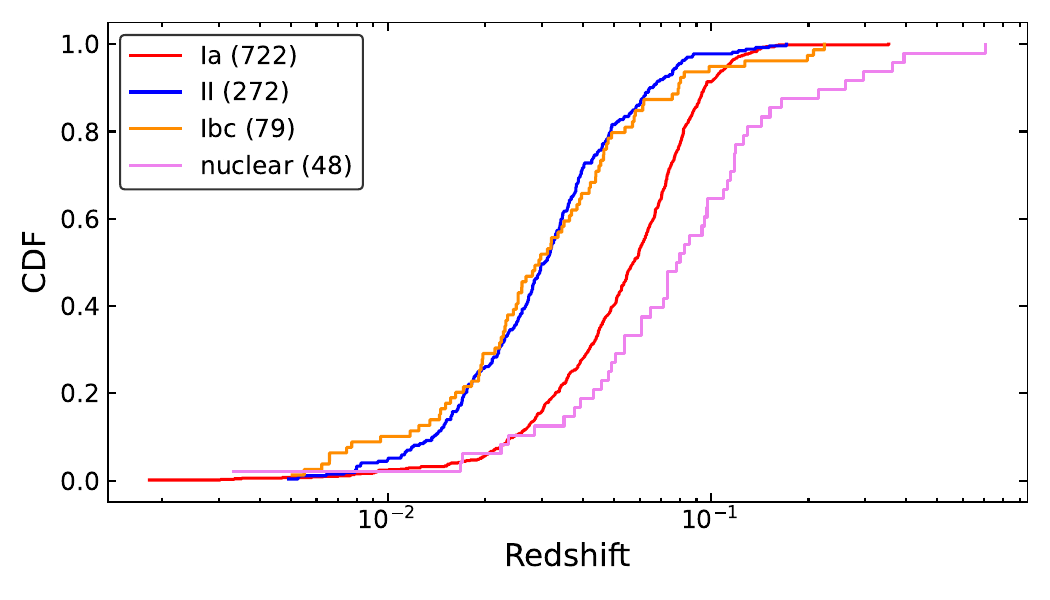}
    \caption{Distribution of redshifts for the extragalactic sources.}
    \label{fig:zdist}
\end{figure}

\begin{figure}
    \centering
    \includegraphics[width=\linewidth]{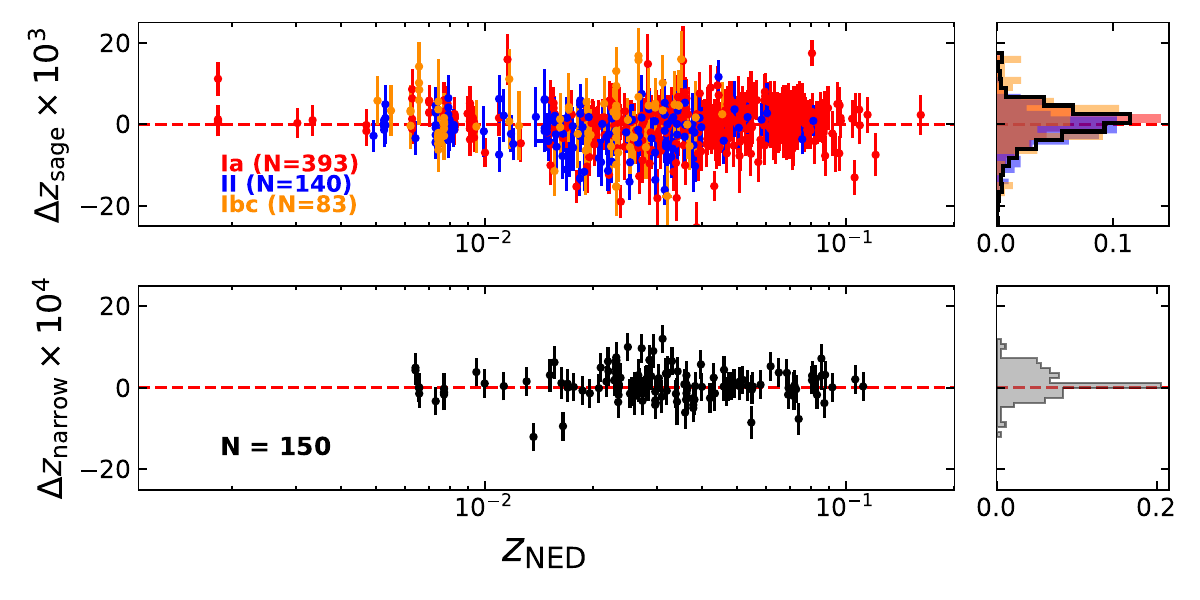}
    \caption{Comparing NED redshifts to those measured from the \sn spectra themselves using \sage (top) and those measured from narrow host-galaxy emission lines (bottom). Note that the y-axes differ by a factor of 10 between the top and bottom panels. The panels along the right side show the collapsed histograms. The top panels are color-coded by \sn type, and the black line in the top right panel is the combined distribution from all \sage redshifts.}
    \label{fig:zcompare}
\end{figure}

After host-galaxy assignment, we query NED to check for existing high-quality redshifts, typically from programs like the 6-degree Field galaxy survey \citep[6dF, ][]{6dF1, 6dF2}, the Sloan Digital Sky Survey \citep[SDSS, ][]{sdss}, and the Dark Energy Spectroscopic Instrument \citep[DESI, ][]{desi1, desi2}. We require spectroscopic (including H~I) redshifts with an uncertainty estimate. We also estimate redshifts from the SN spectra themselves using \sage. The top panel of Fig.~\ref{fig:zcompare} shows that the redshift errors are not strongly correlated with SN type. Based on the observed scatter, we add a systematic error of $\delta z_{\rm sage} = 0.004 \approx 1200~\kms$ in quadrature with the redshift uncertainties reported by \sage so that the residual distribution has unit variance. 

Narrow host-galaxy emission lines in the spectra are also used to measure redshifts. All lines are fit with Gaussian profiles that have the same redshift and velocity FWHM, and the local continuum around each line is modeled as a cubic polynomial. The most common emission lines used are [\ion{O}{2}]$\lambda3727\AAA$, \Ha, [\ion{N}{2}]$\lambda\lambda6548,6584\AAA$, and [\ion{S}{2}]$\lambda\lambda6716,6731\AAA$. Sometimes the fainter \Hg, \Hd, and [\ion{O}{3}]$\lambda4364\AAA$ lines are included. The stronger \Hb and [\ion{O}{3}]$\lambda\lambda4959,5007~\AAA$ lines are used less often because they often fall in the dichroic crossover region ($\approx 5000-5200~\AAA$), which has increased noise and artifacts. To ensure the narrow-line redshifts are robust, we require $\geq2$ lines detected at $\geq 5\sigma$ or $\geq3$ lines detected at $\geq 3\sigma$. The systematic error for narrow-line redshifts is $\delta z = 0.0003\approx 90~\kms$.

Finally, we manually estimate redshifts for spectra showing emission lines that cannot be fit by the narrow-line model, typically broad-line AGN (cf. Fig.~\ref{fig:nuclear}). We use the same emission lines to manually estimate a redshift (excluding Balmer features), and we assign an uncertainty of $\delta z = 0.015 \approx 4500~\kms$ for these cases. Our final distribution of redshifts is shown in Fig.~\ref{fig:zdist}.

\begin{figure*}
    \centering
    \includegraphics[width=0.48\linewidth]{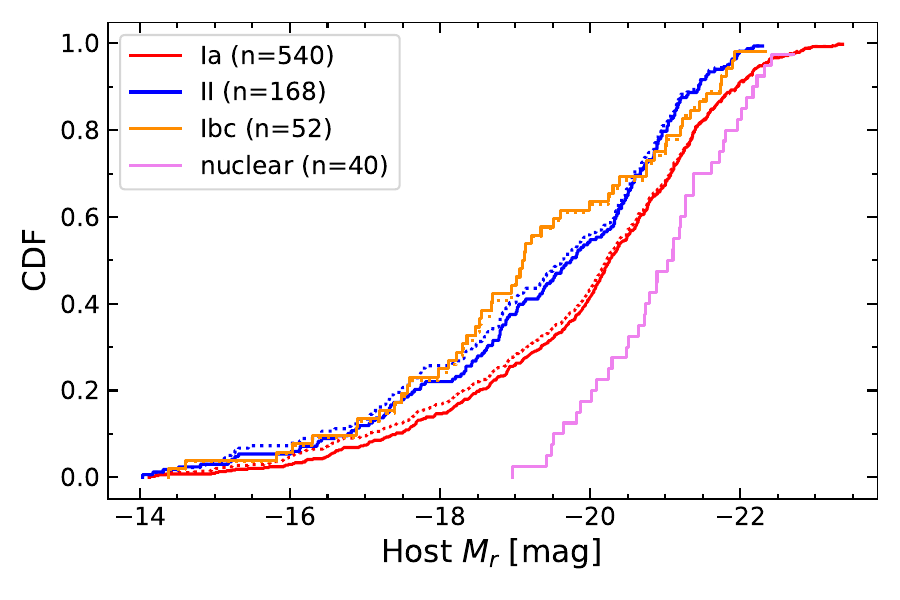}
    \includegraphics[width=0.48\linewidth]{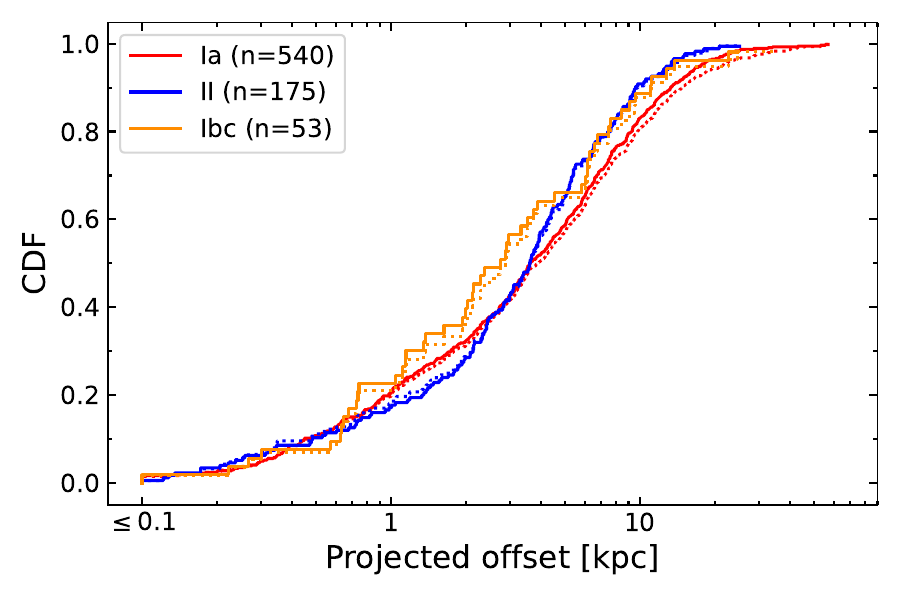}
    
    \caption{\textit{Left}: CDF of absolute $r$-band magnitudes of the host galaxies. \textit{Right}: Projected offsets between SNe and their hosts. Dotted lines show the distributions when tentative hosts are included.}
    \label{fig:hostprops}
\end{figure*}

Next, we estimate distances to each transient. First, we check for redshift-independent distances in the Cosmicflows-4 catalog \citep{cf4} for the most nearby events requiring distance-modulus uncertainties $<0.3$~mag. If a redshift-independent distance is not available, we use the Cosmicflows-4 distance-velocity calculator \citep{valade2024},\input{footnotes/cf4-dv-calc-url} which uses the peculiar-velocity field inferred from the Cosmicflows-4 galaxies to convert an input sky position and redshift ($z\leq 0.1$) into a distance. For the few sources at $z > 0.1$, we use Hubble flow distances adopting the cosmological parameters $H_0 = 70$~\kms and $\Omega_m = 0.3$. Uncertainties in the redshift-based methods add the effect of redshift uncertainty (i.e., re-doing the calculations for $\pm\delta z$) in quadrature with the empirical distance errors measured by \citet{haubner2025}. 

\vspace{0.3cm}
\subsection{Host Photometry and Projected Offsets}\label{subsec:host_phot}

The left panel of Fig.~\ref{fig:hostprops} shows the distribution of absolute $r$-band magnitudes for transient host galaxies in DR1. We choose photometry from, in order of preference, (1) DeCaLS $r$-band magnitudes (automatic shape model, \citealp{dey2019}); (2) Pan-STARRS DR2 $r$-band Kron magnitudes \citep{flewelling2016}; and (3) SDSS DR6 \citep{sdss_dr6} $r$-band (cModel) magnitudes obtained via NED. For the subset of hosts with Pan-STARRS and DeCaLS $r$-band photometry, we measure a median difference of $-0.30$~mag (i.e., DeCaLS is brighter) and a root-mean-square (rms) scatter of $0.17$~mag. We subtract the difference from Pan-STARRS magnitudes to align the survey photometry, and adopt a systematic uncertainty of $\pm0.2$~mag for all host-galaxy photometry.

The adopted host-galaxy coordinates are likely accurate to $\approx 5\arcsec$ in most cases. Many potential hosts in NED originate from the Wide-Field Infrared Explorer (WISE, \citealp{wright2010}) or Galaxy Evolution Explorer (GALEX, \citealp{martin2005}) missions, which have PSFs of $\approx 6\arcsec$. Candidates hosts identified in optical imaging from SDSS, DeCaLS, or Pan-STARRS are often more accurate, closer to $1\arcsec$, but centroiding extended sources starts to dominate the error budget. 
Thus, we suggest a typical host-galaxy astrometric uncertainty of $5\arcsec$ for most cases.

We measure the projected separation between the transient and the host, shown in the right panel of Fig.~\ref{fig:hostprops}. We do not attempt to calculate deprojected distances because we lack positional uncertainties for each source. The \ccsne (II and Ibc) generally track the stellar light of their star-forming hosts, whereas \sneia show a long tail towards larger offsets \citep[e.g., ][]{cronin2021}.

\begin{figure}
    \centering
    \includegraphics[width=\linewidth]{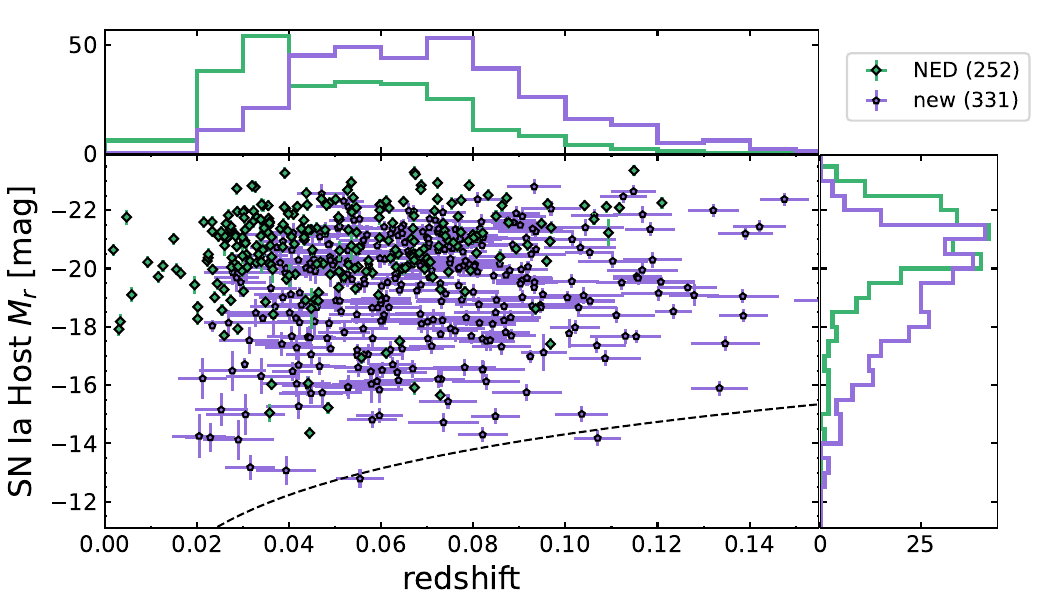}
    \caption{Distribution of redshift and host \Mr for the \sneia in our sample, separated by those with (green) and without (purple) existing spectroscopic redshifts in NED. The dashed line shows a detection limit of $r = 24$~mag.}
    \label{fig:sneia-hostz}
\end{figure}

Fig.~\ref{fig:sneia-hostz} shows the redshift completeness of \snia hosts separated by those with and without existing redshifts from NED. \sneia are used for redshift completeness because they explode in passive and active environments, whereas \ccsne require ongoing or recent star formation \citep[e.g., ][]{ztfbts_paper1}. Luminous galaxies ($M_r \lesssim -20$~mag) within $z\lesssim 0.03$ almost always have a cataloged redshift, and $\gtrsim 50\%$ have a cataloged redshift even at $z\approx 0.1$. Yet galaxies similar to the Magellanic Clouds ($M_r\sim -18~$mag) rarely have prior redshift measurements \citep[e.g., ][]{kulkarni2018}. \sne appear to be a promising avenue for `filling in' the low-luminosity galaxies often overlooked by larger redshift surveys, with more than half of our \snia sample reporting a redshift for the first time. The lower-luminosity \sneii and \sneibc are more likely to be found in nearby galaxies with existing redshifts, but still provide new redshifts for $\sim 150$ ($\sim$40\%) of them.

\section{Light Curves and Phase Estimates}\label{sec:lightcurves}

\begin{figure*}
    \centering
    \includegraphics[width=0.85\linewidth]{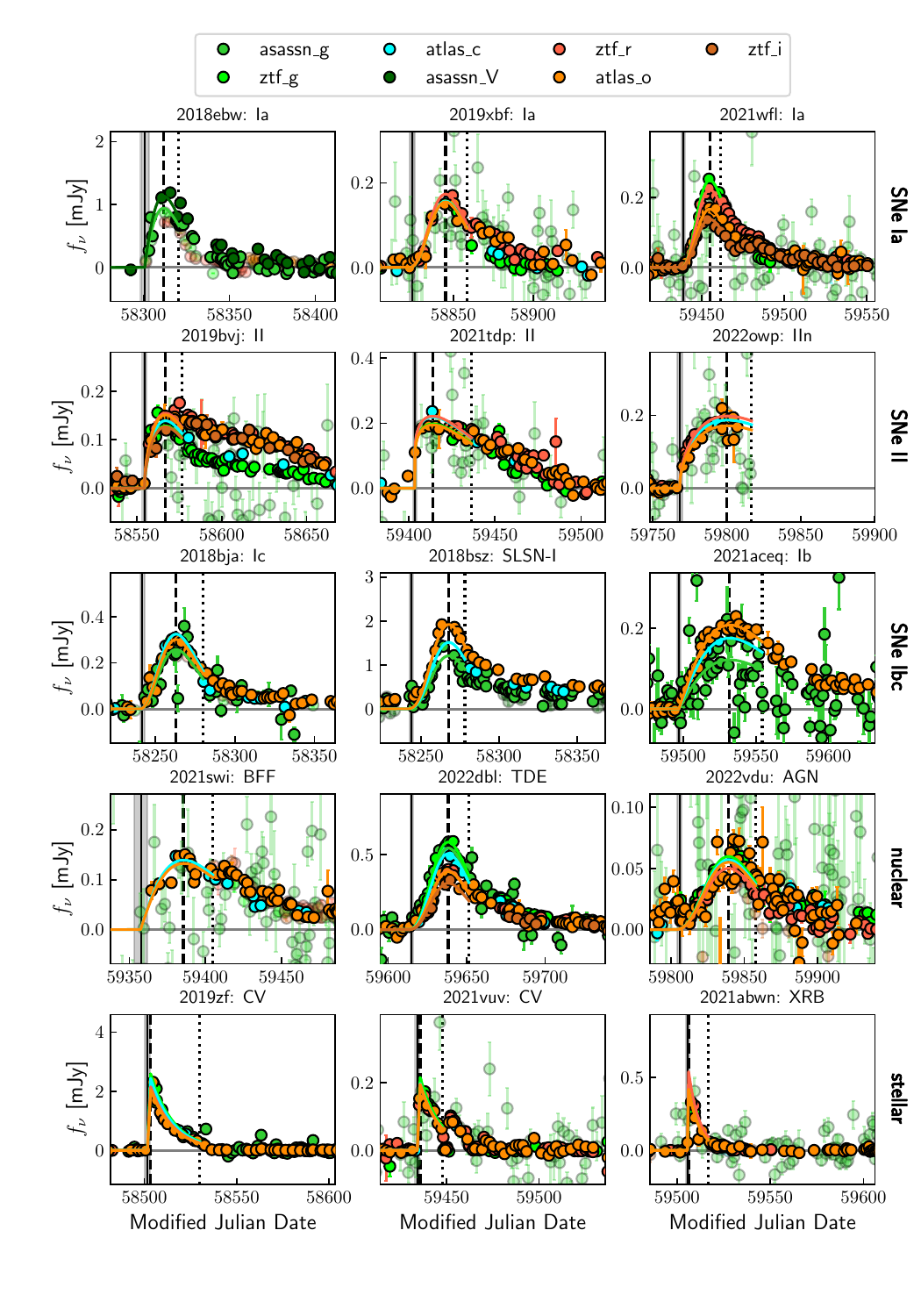}
    \caption{Example light curve fits for 3 randomly-selected events from each spectral class: \sneia (top), \sneii (upper middle), \sneibc (middle), nuclear transients (lower middle), and stellar transients (bottom). Partially-transparent points were omitted when fitting the light curve. The name and subtype are provided above each panel. The vertical solid black line and gray-shaded region show the inferred time of first light $t_1$ and its 84\% confidence interval. The vertical dashed black line shows the measured time of peak flux, and the vertical dotted black line shows the end of the light curve fitting region. All panels span ($t_1 - 20~\rm{d}$) to ($t_{\rm max} + 100~\rm{d}$).}
    \label{fig:lcfit-examples}
\end{figure*}

We compile multi-filter light curves for each of our sources from publicly available survey photometry. We include photometry from ASAS-SN ($g$, $V$), ATLAS ($c$, $o$), and ZTF ($g$, $r$, $i$) queried within $\pm 1$~year of discovery through their respective forced-photometry services \citep{kochanek2017, smith2020, masci2023}. The pre-discovery photometry is used to ensure a correct zero-flux baseline in the image-subtraction measurements, and the post-discovery photometry is used to estimate the reference times needed for accurate phases (i.e., days after explosion or outburst).


\subsection{Phenomenological Light Curve Models}

We use two simple models to estimate the light curve evolution, where we can identify a peak in the light curve near the time of our SNIFS observation(s). The first is the `curved power law' (CPL) model from \citet{vallely2022} with the functional form,

\begin{equation}
    f_\lambda (t) = 
    \begin{cases} 
        b_\lambda & t < t_0, \\
        \Lambda_\lambda\left(\frac{t-t_0}{1+z}\right)^\alpha + b_\lambda& t \geq t_0,
    \end{cases}\label{eq:cpl}
\end{equation}

\noindent for some reference time $t_0$ for a source at redshift $z$ and with a background level $b_\lambda$. The power-law coefficient $\alpha$ is

\begin{equation}
    \alpha \equiv \alpha_1\left(1+\frac{\alpha_2(t-t_0)}{(1+z)}\right),
\end{equation}

\noindent where $\alpha_1$ and $\alpha_2$ are the individual power-law indices. The CPL model is scaled by the factor 

\begin{equation}
    \Lambda_\lambda = A_\lambda\times10^{-\alpha},
\end{equation}
\noindent with $A_\lambda$ denoting the per-filter amplitude of the model, as this helps to break degeneracies between $t_0$ and the power-law indices \citep{miller2020}.

\begin{figure}
    \centering
    \includegraphics[width=\linewidth]{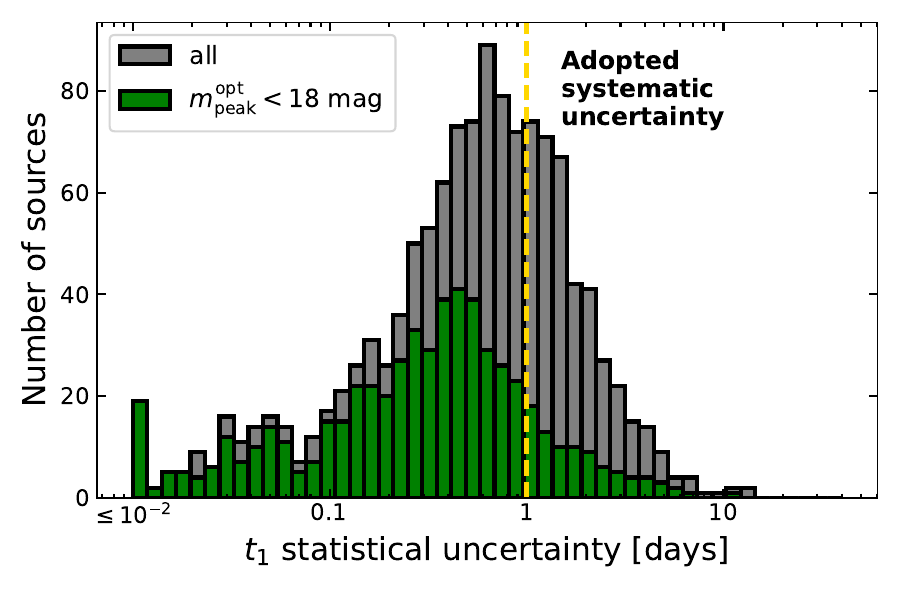}
    \includegraphics[width=\linewidth]{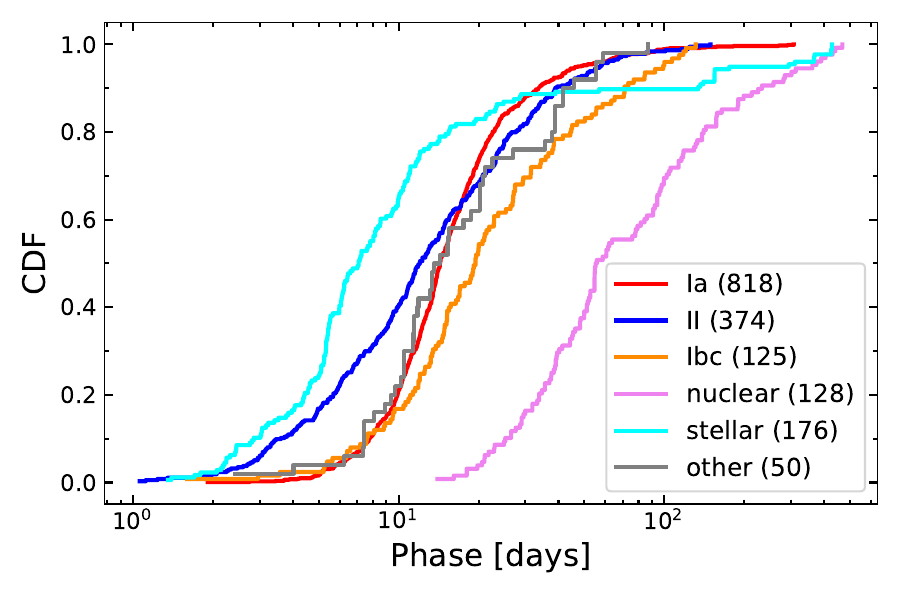}
    \caption{\textit{Top}: The distribution of $t_1$ uncertainties from the light curve fits. The bright subset is shown in green. The vertical dashed yellow line shows our adopted 1-day uncertainty floor. \textit{Bottom}: Distribution of (rest-frame) phases, relative to $t_1$, for the spectra in DR1.}
    \label{fig:dt1_phase_distr}
\end{figure}

Importantly, we assume the power-law indices $\alpha_1$ and $\alpha_2$ are independent of filter, meaning that all filter light curves are simply scaled versions of each other. This assumption of filter-independent light curve shape allows a consistent estimate of $t_0$ that is (mostly) independent of filter sampling and is more robust when the light curve is low SNR or poorly sampled. The background flux level is fit independently for each filter to absorb any offsets in the subtraction baseline. This simple approach enables robust phase estimates for the vast majority of our sample.

The CPL model mainly struggles when the rise is sparsely sampled because $\alpha_1$ becomes poorly constrained. This mostly applies to fast-evolving transients that rise \emph{and} fade within just a few days, typically stellar transients such as CVs, but weather and moon phase also affect photometric sampling. 

We use a `fast rise, exponential decay' (FRED) model to characterize the flux evolution of these fast-evolving sources. This model features a linear rise transitioning to an exponential decline,

\begin{equation}
    f_\lambda(t) = 
    \begin{cases} 
        b_\lambda & t < t_0 \\
        m_\lambda \frac{(t-t_0)}{(1+z)} + b_\lambda & t_0 \leq t < \tmax, \\
        m_\lambda \frac{\trise}{(1+z)} e^{-\beta\frac{(t-t_{\rm max})}{(1+z)}} + b_\lambda & t \geq \tmax,
    \end{cases}\label{eq:fred}
\end{equation}

\noindent with free parameters for the reference time $t_0$, rising slope $m_\lambda$, rise time \trise, and exponential-decay coefficient $\beta$. The time of maximum is $\tmax \equiv t_0 + \trise$. We use the same $t_0$, \trise, and $\beta$ across filters, whereas the rising light curve slope(s) $m_\lambda$ and background level(s) $b_\lambda$ vary per filter.

\begin{figure*}
    \centering
    \includegraphics[width=0.99\linewidth]{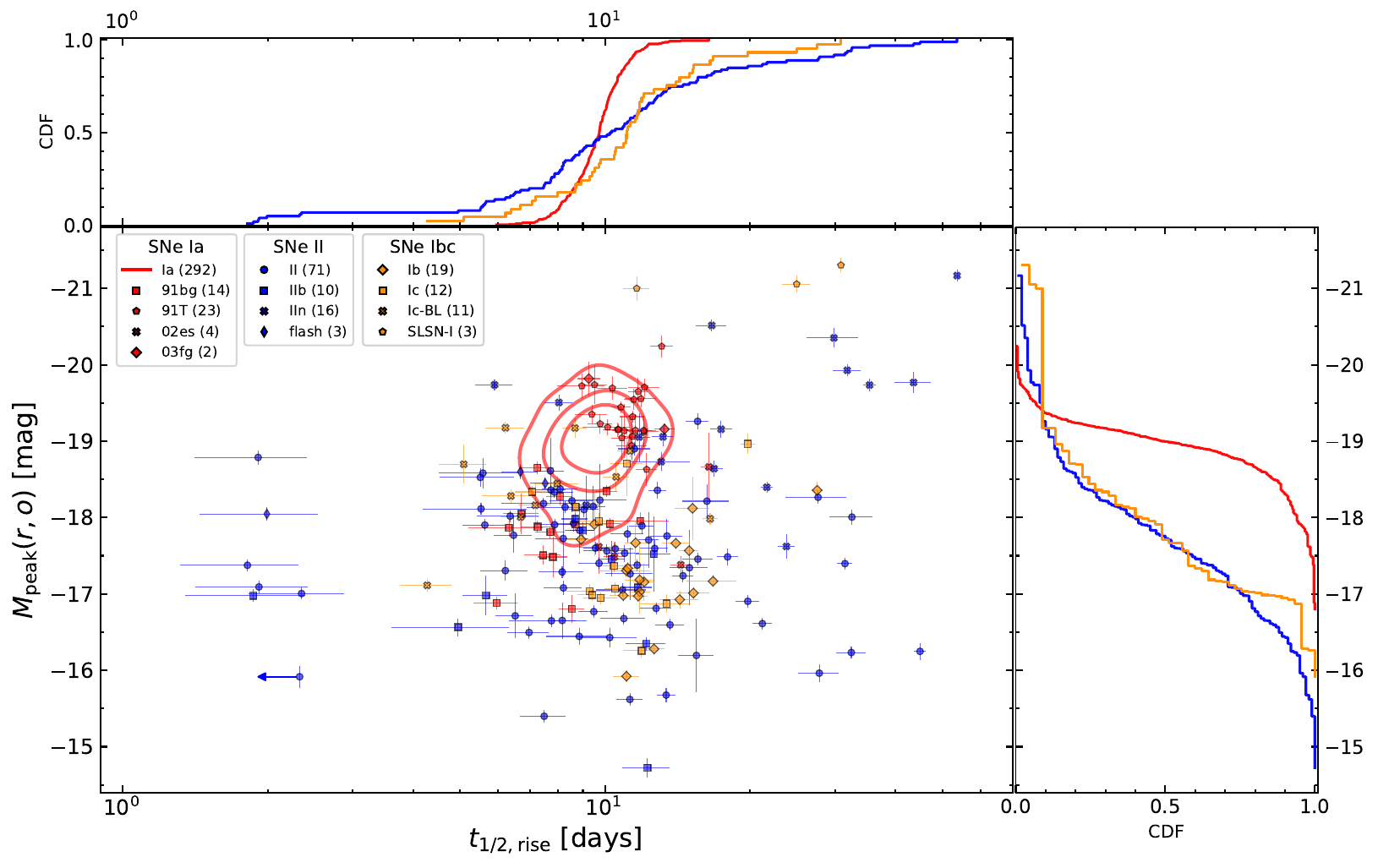}
    \caption{The distribution of \sne in \thalf and $M_{\rm peak}(r,o)$, the averaged peak absolute magnitude in the ATLAS $o$ and ZTF $r$ filters. The peak magnitudes are not corrected for intrinsic host-galaxy extinction, $K$-corrections, or lensing magnification. Only SN (sub-)types with $\geq 2$ sources matching the criteria in \S\ref{subsec:deriv_lc_params} are shown. The top and right panels show the corresponding CDFs along each axis. The 50\%, 75\%, and 95\% density contours are shown for the \sneia, and all other events are shown as points. }
    \label{fig:trise-Mpeak}
\end{figure*}

Both models are fit to the observed multi-filter light curves in flux-density units (mJy) and the models have $2N+3$ degrees of freedom for a source covered by $N$ filters. After an initial least-squares fit, uncertainties are estimated with a Markov Chain Monte Carlo (MCMC) routine using $5000$ steps and 5 walkers per free parameter. The upper and lower bounds imposed on the parameters during MCMC sampling are provided in Table~\ref{tab:mcmc_priors}. By default, light curves are fit up to 10~days after the peak inferred from the initial least-squares fit, but we modify the fitting range as needed to accurately capture the evolution. For sources with complex or multi-peaked light curves, we prioritize fitting the first peak with enough observations to constrain $t_0$.

\input{tables/mcmc_priors}

Fig.~\ref{fig:lcfit-examples} shows a representative set of light curve fits. The models generally provide a faithful representation of the early light curve evolution, adequately describing a wide range of brightness, sampling, and light curve structure. We do not report a light curve fit for a small fraction ($\approx 6\%$) of the sources in DR1. These are most often transients that occurred during Solar conjunction and were only discovered days to weeks later, preventing modeling of the early light curve. We also do not attempt to fit very short outbursts or flares lasting less than 1 day due to insufficient detections. For these sources, all phases are listed relative to the discovery date. 

\subsection{Derived Parameters}\label{subsec:deriv_lc_params}

After fitting each light curve, we derive a set of `secondary' parameters that are the basis for our analysis. These derived parameters are measured directly from the model light curves and their MCMC parameter chains to ensure uncertainties are properly translated to the derived quantities. The derived light curve parameters are (with units in brackets):

\begin{itemize}
    \item $t_1$: the time at which the light curve reaches 1\% of its peak brightness [MJD];
    \item $t_{\rm max}$: the time when the light curve reaches peak brightness [MJD];
    \item $t_{1/2,\rm rise}$: the time it takes the light curve to rise from 50\% of peak flux to peak, corrected for redshift ($t_{\rm max} - t_{50\%})/(1+z)$ [rest-frame days]; 
    \item $f_{\rm max}$: the peak flux [mJy];
    \item $m_{\rm max}$: \fpeak converted to AB magnitudes [mag];
\end{itemize}

\noindent Using these derived quantities reduces some of the biases in the underlying model parameters. For example, the CPL model prefers later $t_0$ and lower $\alpha_1$ when the rising light curve is sparsely sampled or low S/N \citep{miller2020, fausnaugh2023}. Initial testing also showed that the underlying model parameters can depend on which filters are included when fitting the light curve, or how far after peak the light curve is fit. The derived parameters are closer to the data being fit, and thus mainly require only that the models adequately reproduce the observations.

We show the $t_1$ uncertainty distribution and spectral phases in Fig.~\ref{fig:dt1_phase_distr}. Sources without a light curve fit are not included. SCAT prioritizes young/recently-discovered transients, so the phase distribution generally reflects the early brightness evolution. It is easier to obtain early spectra ($\sim$days) of fast-rising events (\sneii, CVs), while the transients powered by the diffusion of radioactive decay energy (\sneia, most \sneibc) require $2-3$ weeks to brighten. The slowest transients are accretion-powered nuclear sources which often take weeks to months to rise to peak brightness.

Fig.~\ref{fig:trise-Mpeak} shows the absolute peak magnitude versus \thalf for the different SN (sub)types. We use $M_{\rm max}(r,o)$, the absolute magnitude averaged over the ZTF $r$- and ATLAS $o$-band filters. Using $M_{\rm max}(g,c)$ produces similar results, but is more affected by host reddening. The ATLAS $o$-band data provide the most reliable combination of cadence and depth for the majority of sources. The ZTF $i$-band is excluded because it is often poorly sampled and typically evolves noticeably differently from the other filters. We apply (rather conservative) quality cuts to remove poorly sampled or spurious light curves: \mpeak $\leq 19.5$~mag in any filter, $\chi^2_\nu \leq 2$, and $\geq 3$ photometric detections ($\geq 3\sigma$) during the rise ($t_1 < t < \tmax$). For sources with $\thalf \leq 10$~days, we require an uncertainty on \thalf of $< 1.5$~days. An absolute cut on \thalf uncertainty penalizes slowly-rising sources, so we switch to a cut on the fractional \thalf uncertainty of $\leq 0.15$ for sources with $\thalf > 10$~d so the two approaches agree at the midpoint. Using a hard cut on \thalf simply removes some of the slower \sneii and \sneiin.

The different SN types separate relatively well in this space, tracing the different physical processes that govern their rising light curves \citep[e.g., ][]{arnett1982, khatemi2024}. The \sneia show a correlation between \thalf and peak brightness \citep[e.g., ][]{Hayden2010}, similar to the original `Phillips relation' used for \snia standardization \citep{phillips1993}, although it is partially obscured by the broad distribution of \sneii rise-times.
For \sneii, the rising light curve is typically tracing the shock-cooling of the H envelope, which depends on the progenitor radius, the explosion energy, and the presence of nearby CSM \citep[e.g., ][]{morag2024}. The \sneiin typically take longer to rise to a more luminous peak than `standard' \sneii, reflecting the conversion of kinetic energy into photons by the CSM shock. The \sneibc generally evolve on similar timescales as \sneia but reach lower luminosities due to their comparatively lower radioactive $^{56}$Ni yields. The \sneibc show increased scatter from the complexities of massive-star evolution, tracing differences in pre-explosion mass loss, rotation, binarity, and He core mass \citep[e.g., ][]{sukhbold2016}.

Despite our best efforts, there are still situations where the light curve fitting struggles to produce reliable results. Strong color evolution in bright transients (i.e., S/N is not the limiting factor) is the most common failure mode, typically \sneibc that rapidly redden as they approach peak brightness \citep[e.g., ][]{zheng2022}. The MCMC chains get stuck in local minima, producing unreasonably low phase uncertainties ($\lesssim 15$~minutes). We add generous 1-day and 10\% (0.1~mag) systematic uncertainties in quadrature with the statistical uncertainties for the temporal and flux (magnitude) measurements, respectively, to prevent over-interpretation of the light curve fits.

\section{Summary}\label{sec:summary}

\begin{figure*}
    \centering
    \includegraphics[width=0.99\linewidth]{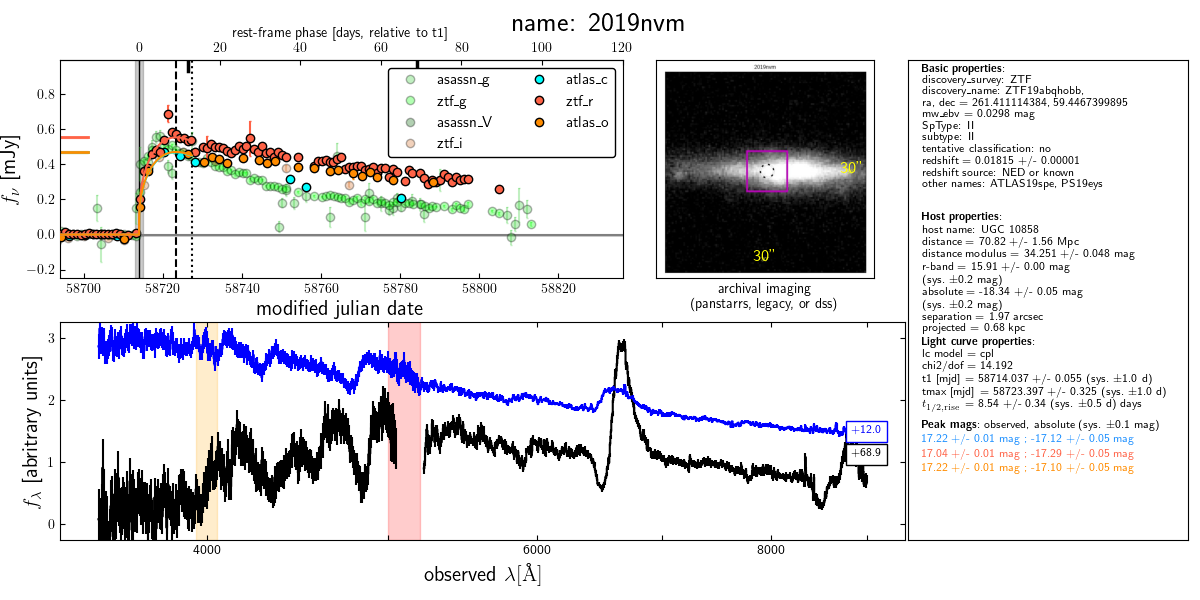}
    \caption{Example `summary' plot for the \snii 2019nvm. \textit{Top left}: Survey light curves and fits (cf. Fig.~\ref{fig:lcfit-examples}). Partially-transparent points were not included in the light curve fit. Ticks along the top axis denote spectra epochs. Horizontal ticks along the left axis represent the filter-specific peak fluxes. \textit{Top middle}: Archival imaging ($30\arcsec\times30\arcsec$) of the transient location from Pan-STARRS, DeCaLS, or DSS. The magenta square shows the approximate SNIFS FoV (cf. Fig.~\ref{fig:quality_examples}). \textit{Bottom left}: Spectroscopic time series, with phases relative to $t_1$ shown along the right side of the plot. The orange and red shaded regions show the typical wavelengths affected by poor dichroic corrections (cf. Fig.~\ref{fig:raster}). \textit{Right}: Aggregated properties for the transient, including base properties (names, coordinates, spectral types, redshift), host-galaxy information, and the results of the light curve fit including \thalf and peak magnitudes.}
    \label{fig:example_summary}
\end{figure*}

We have presented an overview of SCAT DR1 containing \numSpectra spectra of \numSources transient sources collected between March 2018 -- January 2023, including \nspexIa spectra of \numIa \sneia, \nspexII spectra of \numII \sneii, \nspexIbc spectra of \numIbc \sneibc, \nspexNuclear spectra of \numNuclear nuclear transients, \nspexStellar spectra of \numStellar stellar phenomena, and \nspexOther spectra of \numOther objects in the `other' category.  We augment the spectra with host-galaxy associations, redshifts, distances, survey light curves, phase estimates, and peak-brightness measurements. 

DR1 contains the extracted 1D SNIFS spectra in \textsc{fits} and \textsc{ascii} formats, the combined multi-survey light curve, the light curve model fit and parameters, and a $30\arcsec\times30\arcsec$ image of the local environment. These products are combined into a `summary' plot as shown in Fig.~\ref{fig:example_summary}. The data are available from a Zenodo repository\input{footnotes/zenodo} which includes detailed documentation of the DR1 structure and file contents. Most users will be interested in the science data products ($\approx 1$~GB) but we also provide smaller packages with just the summary figures for each source ($\approx 200$~MB) and just the \textsc{ascii}-formatted spectra ($\approx 35$~MB), which should be the default spectra for most science cases. The standard-star spectra used for calibration are also included in a separate package ($\approx 300$~MB). A brief description of the final summary tables is provided in Appendix~\ref{app:products}.

SCAT is an ongoing project, and we expect future data releases to include both new observations obtained since DR1 and improved extraction and calibration methods. SCAT data quality has improved over time thanks to several upgrades to the UH~2.2m telescope, the SNIFS instrument, and our data processing pipelines. The UH2.2m is currently transitioning to robotic operations, enabling fully-automated transient classification, an easily replicable selection function, and fast-turnaround spectra of very young ($\sim$hours) transients. There are several avenues of ongoing pipeline improvement, such as better dichroic corrections, scene modeling to reduce host contamination, and absolute spectrophotometry.
\newline

\vspace{0.5cm}

\textbf{Facilities}: UH2.2m (SNIFS)

\vspace{0.5cm}
\textbf{Software}: astropy \citep{astropy1, astropy2}; numpy \citep{numpy1, numpy2}; matplotlib \citep{matplotlib}; lmfit \citep{lmfit}; scipy \citep{scipy}; spectres \citep{spectres}; emcee \citep{emcee}; pandas \citep{pandas}

\section*{Acknowledgments}

We thank Chris Kochanek and Kris Stanek for useful discussions about the project, and thank Klaus Hodapp, Greg Aldering, John Tonry, and Brent Tully for contributing UH2.2m time.  

M.A.T. acknowledges support from the National Science Foundation through grants AST-2307385 and AST-2407206, and the Gordon and Betty Moore Foundation through awards GBMF5490 and GBMF10501.

W.B.H. acknowledges support from the National Science Foundation Graduate Research Fellowship Program under Grant Nos.\ 1842402 and 2236415. Any opinions, findings, conclusions, or recommendations expressed in this material are those of the authors and do not necessarily reflect the views of the National Science Foundation.

J.T.H. acknowledges support from NASA through the NASA Hubble Fellowship grant HST-HF2-51577.001-A, awarded by STScI. STScI is operated by the Association of Universities for Research in Astronomy, Incorporated, under NASA contract NAS5-26555.

C.A. acknowledges support from NASA grants JWST-GO-02114, JWST-GO-02122, JWST-GO-03726, JWST-GO-04217, JWST-GO-04436, JWST-GO-04522, JWST-GO-05057, JWST-GO-05290, JWST-GO-06023, JWST-GO-06213, JWST-GO-06583, and JWST-GO-06677. Support for these programs was provided by NASA through a grant from the Space Telescope Science Institute, which is operated by the Association of Universities for Research in Astronomy, Inc., under NASA contract NAS5-03127.

CRA is supported by the European Research Council (ERC) under the European Union's Horizon 2020 research and innovation program (grant agreement No. 948381, PI: M. Nicholl).

\appendix

\section{Example Error Model for EG131}\label{app:errmodel}

To derive the error model described in \S\ref{subsec:errfloor} and shown in Fig.~\ref{fig:error_floor}, we start with all spectrophotometric standard-star exposures. We only use the white dwarf standards EG131, GD71, GD153, Feige 67, Geige 110, and G191-B2B because their spectra are nearly featureless. Including the other spectrophotometric standard stars yields similar results, except for increased errors around the Balmer lines due to spectral resolution mismatches.

\begin{figure}
    \centering
    \includegraphics[width=0.9\linewidth]{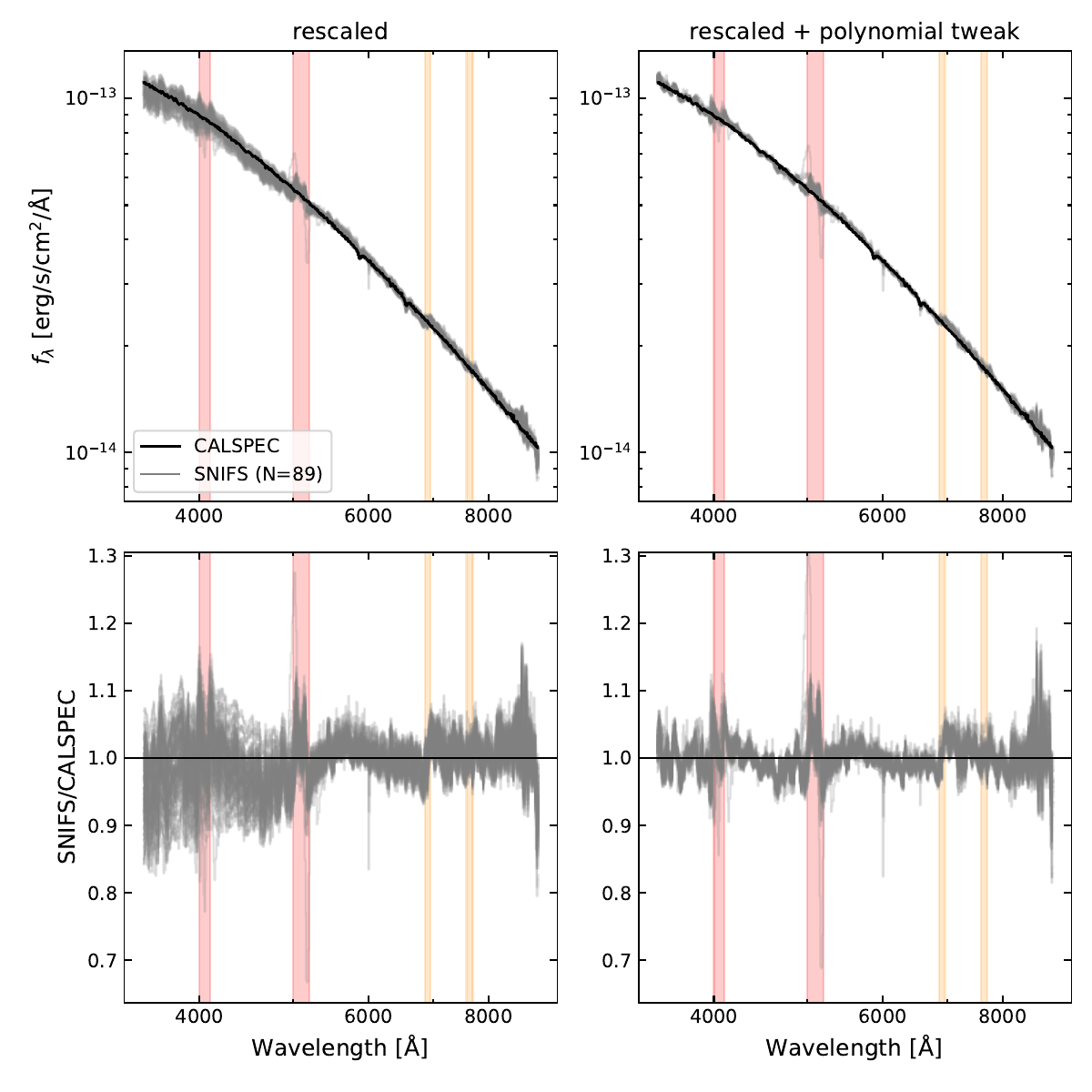}
    \caption{Example of the flux re-normalization process used to derive the empirical error model. The top left panel shows the rescaled SNIFS spectra (gray) of EG131 compared to the CALSPEC spectrum (black). The bottom left panel shows the ratio of these as a function of wavelength. The right panels show the same thing, except with a 7th-order polynomial correction applied to reduce broadband errors in the flux and atmospheric corrections. Red and orange shaded regions represent wavelengths affected by variable dichroic throughput and strong telluric bands, respectively.}
    \label{fig:eg131_errmodel}
\end{figure}

Fig.~\ref{fig:eg131_errmodel} shows an example of the flux re-normalization process for the well-observed standard EG131. We do not perform spectrophotometry in this DR, so the spectra are first rescaled to match the median flux of the CALSPEC spectrum. These rescaled spectra are shown in the top-left panel, and the ratio when compared to CALSPEC is shown in the lower-left panel. The spectral slopes are reliable to a few percent across the full spectral range, reaching an rms of $\approx 5\%$ at bluer wavelengths. The largest source of error is the inter-channel scaling because of the variable dichroic throughput. We prioritize a smooth transition between the B and R channels to ease the analysis of science spectra (e.g., \textsc{snid-sage} interprets any sharp transitions between channels as potential SN features). This results in the broadband spectral slope inheriting any errors from incorrect inter-channel scaling. Variations in atmospheric throughput account for a few percent of the variations \citep[e.g., ][]{buton2013, rubin2022}.

The empirical error model is designed to capture pixel-to-pixel variations introduced by our reduction and calibration procedures, especially around the dichroic and telluric regions. We use a simple 7th-order polynomial model to minimize the effects of broadband flux calibration errors, as shown in the right-hand panels of Fig.~\ref{fig:eg131_errmodel}. The pixel-to-pixel variations generally decrease to a few percent, as shown in Fig.~\ref{fig:error_floor}, except for wavelengths affected by strong dichroic or telluric features where calibration errors can reach $\approx 10\%$ in extreme cases. The remaining large-scale ($\approx 1000~\AAA$) structure in the lower-right panel of Fig.~\ref{fig:eg131_errmodel} is largely due to errors in the flux calibration curves from poor dichroic or telluric corrections, which the polynomial fit tries to account for by adjusting the flux response of adjacent regions. 

\section{Brief Summary of Data Tables}\label{app:products}

DR1 provides 2 `master tables' that aggregate most of the final quantities. The first is the master source table (one entry per object) that includes the following columns:
\input{sourcetable_columns}
\vspace{0.2cm}
For quantities with an uncertainty estimate, the column names have \lst{_error} appended to the end. If the uncertainty includes a systematic error term, we provide the statistical contribution with \lst{_error_stat} appended to the column name.

\vspace{0.2cm}

The second table is the spectrum table (one entry per 1D spectrum) that includes the following columns (excluding overlaps with the master source table):
\input{spectable_columns}

\vspace{0.5cm}
\noindent The documentation in the Zeonodo repository\input{footnotes/zenodo} is the final authority on DR1 data products and structure.

\bibliography{ref}{}
\bibliographystyle{aasjournal}



\end{document}

%% file: defs.tex
\usepackage{xspace, color, orcidlink}
\usepackage{amsmath}
\usepackage{listings}
\newcommand{\lst}[1]{\lstinline{#1}\xspace}
\newcommand{\TABLEFONTSIZE}{\footnotesize}
\newcommand{\NORMALFONTSIZE}{\normalsize}
\newcommand{\padspace}{\vspace{0.2cm}}

\usepackage{hyperref}
\hypersetup{
    colorlinks=true,
    linkcolor=blue,
    citecolor=blue,
    urlcolor=blue,
    }

\newcommand{\um}{\hbox{\ensuremath{\mu \rm m}}\xspace}
\newcommand{\kms}{\hbox{\ensuremath{\rm{km}~\rm s^{-1}}}\xspace}

\newcommand{\AAA}{\hbox{\ensuremath{\rm \AA}}\xspace}

\newcommand{\Ha}{\hbox{H$\alpha$}\xspace}

\newcommand{\ccsne}{CC~SNe\xspace}
\newcommand{\ccsn}{CC~SN\xspace}
\newcommand{\sn}{SN\xspace}
\newcommand{\sne}{SNe\xspace}
\newcommand{\sneia}{SNe Ia\xspace}
\newcommand{\snia}{SN Ia\xspace}
\newcommand{\sneii}{SNe II\xspace}
\newcommand{\snii}{SN II\xspace}

\newcommand{\sneib}{SNe~Ib\xspace}

\newcommand{\sneic}{SNe~Ic\xspace}

\newcommand{\sneibc}{SNe Ibc\xspace}

\newcommand{\sneiin}{SNe~IIn\xspace}

\newcommand{\slsnei}{SLSNe-I\xspace}

\newcommand{\sage}{\textsc{snid-sage}\xspace}
\newcommand{\Hb}{\hbox{\ensuremath{\rm H \beta}}\xspace}
\newcommand{\Hg}{\hbox{\ensuremath{\rm H \gamma}}\xspace}
\newcommand{\Hd}{\hbox{\ensuremath{\rm H \delta}}\xspace}

\newcommand{\trise}{\hbox{\ensuremath{t_{\rm rise}}}\xspace}
\newcommand{\tmax}{\hbox{\ensuremath{t_{\rm max}}}\xspace}

\newcommand{\thalf}{\hbox{\ensuremath{t_{1/2,\rm rise}}}\xspace}
\newcommand{\fpeak}{\hbox{\ensuremath{f_{\rm max}}}\xspace}
\newcommand{\mpeak}{\hbox{\ensuremath{m_{\rm max}}}\xspace}
\newcommand{\HeII}{\ion{He}{2}\xspace}

\newcommand{\prost}{\textsc{Prost}\xspace}

\newcommand{\Mr}{\hbox{\ensuremath{M_r}}\xspace}

\input{counts}

%% file: counts.tex
\newcommand{\numSources}{1331\xspace}
\newcommand{\numSpectra}{1812\xspace}
\newcommand{\numIa}{722\xspace}
\newcommand{\nspexIa}{838\xspace}
\newcommand{\numII}{271\xspace}
\newcommand{\nspexII}{388\xspace}
\newcommand{\numIbc}{79\xspace}
\newcommand{\nspexIbc}{126\xspace}
\newcommand{\numNuclear}{48\xspace}
\newcommand{\nspexNuclear}{172\xspace}
\newcommand{\numStellar}{173\xspace}
\newcommand{\nspexStellar}{230\xspace}
\newcommand{\numOther}{38\xspace}
\newcommand{\nspexOther}{58\xspace}

%% file: AUTHORS2.tex
\author{\vspace{-1.3cm}
    Michael~A.~Tucker$^{1,2,\ast}$\orcidlink{0000-0002-2471-8442}}
\author{Mark~E.~Huber$^{3}$\orcidlink{0000-0003-1059-9603}}
\author{Benjamin~J.~Shappee$^{3}$\orcidlink{0000-0003-4631-1149}}
\author{Jason~T.~Hinkle$^{4,5,\dagger}$\orcidlink{0000-0001-9668-2920}}
\author{Willem~B.~Hoogendam$^{3,\ddagger}$\orcidlink{0000-0003-3953-9532}}
\author{Charlotte~R.~Angus$^{6}$\orcidlink{0000-0002-4269-7999}}
\author{Chris~Ashall$^{3}$\orcidlink{0000-0002-5221-7557}}
\author{Katie~Auchettl$^{7,8}$\orcidlink{0000-0002-4449-9152}}
\author{Kenneth~C.~Chambers$^{3}$\orcidlink{0000-0001-6965-7789}}
\author{Dhvanil~D.~Desai$^{3}$\orcidlink{0000-0002-2164-859X}}
\author{Aaron~Do$^{9}$\orcidlink{0000-0003-3429-7845}}
\author{Joseph~Ghammashi$^{2}$\orcidlink{0009-0008-2932-8030}}
\author{Catherine~J.~Grier$^{10}$\orcidlink{0000-0001-9920-6057}}
\author{Joanna~Herman$^{3}$}
\author{Thomas~de~Jaeger$^{11}$\orcidlink{0000-0001-6069-1139}}
\author{Jodie~Kiyokawa$^{10}$\orcidlink{0009-0003-0398-0382}}
\author{Miranda~Y.~Kong$^{3}$\orcidlink{0009-0005-5121-2884}}
\author{Thomas~B.~Lowe$^{3}$\orcidlink{0000-0002-9438-3617}}
\author{Eugene~A.~Magnier$^{3}$\orcidlink{0000-0002-7965-2815}}
\author{Anna~V.~Payne$^{12}$\orcidlink{0000-0003-3490-3243}}
\author{Sara~Romagnoli$^{7}$\orcidlink{0009-0003-8153-9576}}
\author{David~Rubin$^{13,14}$\orcidlink{0000-0001-5402-4647}\vspace{0.3cm}}

\affiliation{$^{1}$ Center for Cosmology and AstroParticle Physics, 191 W Woodruff Ave, Columbus, OH 43210}
\affiliation{$^{2}$ Department of Astronomy, The Ohio State University, 140 W 18th Ave, Columbus, OH 43210}
\affiliation{$^{3}$ Institute for Astronomy, University of Hawai`i, Honolulu, HI 96822}
\affiliation{$^{4}$ Department of Astronomy, University of Illinois Urbana-Champaign,\\1002 West Green Street, Urbana, IL 61801, USA}
\affiliation{$^{5}$ NSF-Simons AI Institute for the Sky (SkAI), 172 E. Chestnut St., Chicago, IL 60611, USA}
\affiliation{$^{6}$ Astrophysics Research Centre, School of Mathematics and Physics,\\Queen’s University Belfast, Belfast BT7 1NN, UK}
\affiliation{$^{7}$ OzGrav, School of Physics, The University of Melbourne, Parkville, VIC, Australia}
\affiliation{$^{8}$ Department of Astronomy and Astrophysics, University of California, Santa Cruz, CA, USA}
\affiliation{$^{9}$ Institute of Astronomy and Kavli Institute for Cosmology, Madingley Road, Cambridge CB3 0HA, UK}
\affiliation{$^{10}$ Department of Astronomy, University of Wisconsin-Madison, Madison, WI 53706, USA}
\affiliation{$^{11}$ LPNHE, (CNRS/IN2P3, Sorbonne Universit\'{e}, Universit\'{e} Paris Cit\'{e}), Laboratoire de Physique Nucl\'{e}aire et de Hautes \'{E}nergies, 75005, Paris, France}
\affiliation{$^{12}$ Space Telescope Science Institute, 3700 San Martin Drive, Baltimore, MD 21218, USA}
\affiliation{$^{13}$ Department of Physics and Astronomy, University of Hawai`i at M{\=a}noa, Honolulu, Hawai`i 96822}
\affiliation{$^{14}$ Physics Division, E.O. Lawrence Berkeley National Laboratory,\\1 Cyclotron Rd., Berkeley, CA, 94720, USA}

\altaffiliation{$^{\ast}$ CCAPP Fellow}
\altaffiliation{$^{\dagger}$ NHFP Einstein Fellow}
\altaffiliation{$^{\ddagger}$ NSF Fellow}
\email{Contact: tuckerma95@gmail.com}

%% file: footnotes/zenodo.tex
\footnote{\url{https://zenodo.org/records/19188201}}

%% file: footnotes/tns.tex
\footnote{\url{https://www.wis-tns.org/}}

%% file: footnotes/Rch_H2O.tex
\footnote{The SNIFS R channel reaches $1\um$ but we omit data beyond 9200~\AAA because it is contaminated by a strong H$_2$O telluric feature.}

%% file: footnotes/multispec.tex
\footnote{Counting the number of nights with spectra, not the total number of spectra.}

%% file: footnotes/snifs.tex
\footnote{The more sophisticated SNIFS reduction pipeline developed by the SNFactory for \snia cosmology reaches a median per-exposure repeatability of $\approx 1.4\%$ \citep{rubin2022}.}

%% file: footnotes/reviewers.tex
\footnote{The following co-authors assisted with the data review: CRA, KA, DDD, AD, JG, JTH, WBH, JK, SR, MAT}

%% file: tables/object_counts.tex
\begin{table}
\centering
\caption{Number of objects and spectra by spectral type and subtype.}\label{tab:object_counts}
\begin{tabular}{ccrr}
\hline\hline
SpType & Subtype & $N_{\rm obj}$ & $N_{\rm spex}$ \\ 
\hline
\textbf{Ia} & & \textbf{722} & \textbf{838} \\ 
 & Ia & 641 & 708 \\ 
 & 91T & 40 & 45 \\ 
 & 91bg & 21 & 37 \\ 
 & 02es & 8 & 11 \\ 
 & 03fg & 6 & 12 \\ 
 & 02cx & 3 & 8 \\ 
 & CSM & 2 & 3 \\ 
 & lensed & 1 & 14 \\ 
\hline
\textbf{II} & & \textbf{271} & \textbf{388} \\ 
 & II & 180 & 236 \\ 
 & IIn & 48 & 68 \\ 
 & IIb & 32 & 54 \\ 
 & flash & 11 & 30 \\ 
\hline
\textbf{Ibc} & & \textbf{79} & \textbf{126} \\ 
 & Ib & 30 & 41 \\ 
 & Ic & 20 & 34 \\ 
 & Ic-BL & 17 & 19 \\ 
 & SLSN-I & 7 & 25 \\ 
 & Ibn & 4 & 5 \\ 
 & Icn & 1 & 2 \\ 
\hline
\textbf{nuclear} & & \textbf{48} & \textbf{172} \\ 
 & AGN & 26 & 29 \\ 
 & TDE & 15 & 86 \\ 
 & ANT & 5 & 55 \\ 
 & BFF & 2 & 2 \\ 
\hline
\textbf{stellar} & & \textbf{173} & \textbf{230} \\ 
 & CV & 122 & 130 \\ 
 & XRB & 13 & 22 \\ 
 & Mdwarf & 10 & 10 \\ 
 & YSO & 8 & 26 \\ 
 & nova & 8 & 11 \\ 
 & other & 5 & 5 \\ 
 & LBV & 4 & 4 \\ 
 & LRN & 2 & 3 \\ 
 & dipper & 1 & 19 \\ 
\hline
\textbf{other} & & \textbf{38} & \textbf{58} \\ 
 & SN & 20 & 21 \\ 
 & unknown & 13 & 22 \\ 
 & BLLac & 3 & 3 \\ 
 & FBOT & 1 & 11 \\ 
 & ILRT & 1 & 1 \\ 
\hline
\hline
\end{tabular}
\end{table}

%% file: footnotes/mdwarf.tex
\footnote{We only look for the molecular bands associated with low-mass stars and do not attempt to distinguish between M~dwarfs and K~dwarfs.}

%% file: footnotes/cow.tex
\footnote{The SNIFS observations of AT~2018cow included in DR1 were originally published by \citet{cow1}.}

%% file: footnotes/cf4-dv-calc-url.tex
\footnote{\url{https://edd.ifa.hawaii.edu/CF4calculator/}}

%% file: tables/mcmc_priors.tex
\begin{table}[]
    \centering
    \caption{Upper and lower bounds for the flat (uninformative) priors used by each model during MCMC sampling. $\star$ denotes results from the initial least-squares fit. $t_d$ is the discovery date reported to TNS.}
    \label{tab:mcmc_priors}

    \begin{tabular}{lcc}
    \hline\hline
    Parameter & Lower bound & Upper bound \\
    \hline
    & \underline{CPL model} \\
    $\alpha_1$ & $10^{-3}$ & 3.5 \\
    $\alpha_2$ [1/d] & $-1$ & $-10^{-3}$ \\
    $A_\lambda$ [mJy] & 0 & $10^5$ \\
    \hline 
    & \underline{FRED model} \\
    $m_\lambda$ [mJy/d] & $10^{-5}$ & 10 \\
    $\beta$ & $10^{-3}$ & 10 \\
    $t_{\rm rise}$ [d] & 0.1 & 50 \\ 
    \hline 
    & \underline{Both models} \\
    $t_0$ [MJD] & $\rm{min}(t_0^{\rm \star} - 10\sigma_{t0}^{\rm \star}, t_{\rm d} - 10)$ & $\rm{min}(t_0^\star+5\sigma_{t0}^\star, t_{\rm d})$  \\
    $b_\lambda$ [mJy] & $-1$ & 1 \\        
    \hline\hline
    \end{tabular}
\end{table}

%% file: sourcetable_columns.tex
\TABLEFONTSIZE
\begin{itemize}
    \item \lst{tns_name}: IAU/TNS name of the object
    \item \lst{ra}: Right Ascension from TNS [deg];
    \item \lst{dec}: Declination from TNS [deg];
    \item \lst{discovery_survey}: Discovering survey (who reported it to TNS), if applicable
    \item \lst{discovery_name}: name assigned by the discovering survey, if present
    \item \lst{other_names}: alternative survey names reported to TNS
    \item \lst{discovery_date}: date of discovery, from TNS
    \item \lst{discovery_mjd}: \lst{discovery_date} converted to Modified Julian Date (MJD)
    \item \lst{num_snifs}: number of SNIFS spectra released in DR1
    \item \lst{mw_ebv}: Milky Way reddening $E(B-V)$ along the line-of-sight [mag]
    \item \lst{sptype}: Assigned spectral type, one of: SN Ia, SN II, SN Ibc, nuclear, stellar, or other;
    \item \lst{subtype}: Assigned spectral sub-type;
    \item \lst{tentative_classification}: `true' if SpType is considered uncertain;
    \item \lst{tns_sptype}: The spectral type listed on TNS as of writing;
    \item \lst{tns_redshift}: The redshift listed on TNS at the time of writing;
    \item \lst{redshift}: Adopted heliocentric redshift;
    \item \lst{redshift_source}: How the redshift was derived, one of: `NED or known', `narrow host lines', `snid-sage', or `manual';
    \item \lst{distance}: Adopted distance [Mpc];
    \item \lst{distmod}: Distance modulus [mag];
    \item \lst{distance_method}: How the distance was derived, one of: `CF4' for the Cosmicflows-4 catalog, `local\_flow' for the Cosmicflows-4 distance-velocity calculator, `hubble\_flow', or `gaia'
    \item \lst{references}: Bibcodes for associated references including discovery, classification, redshifts, distances, and previous publications
\end{itemize}
\NORMALFONTSIZE

\padspace
If a host galaxy was successfully matched to the source, the following fields are populated:

\TABLEFONTSIZE
\begin{itemize}
    \item \lst{host_name}: The name of the galaxy, either from NED or one of the \textsc{Prost} input catalogs
    \item \lst{tentative_host}: `True' if the host association is considered tentative
    \item \lst{host_ra}: Right ascension of the galaxy [deg]
    \item \lst{host_dec}: Declination of the galaxy [deg]
    \item \lst{host_sep_arcsec}: Measured offset between the host and transient coordinates [arcsec]
    \item \lst{host_sep_kpc}: Projected distance between the transient and the galaxy [kpc]
    \item \lst{host_rmag}: $r$-band magnitude of the galaxy, if available [mag]
    \item \lst{host_rmag_source}: Source for the r-band photometry, either `panstarrs', `decals', or `sdss'
    \item \lst{host_absrmag}: Absolute $r$-band magnitude of the galaxy [mag]
\end{itemize}
\NORMALFONTSIZE
\padspace

\noindent The next entry signals if one of the two light curve models were fit to the survey photometry, \\\\
\lst{lc_model}: light curve model used, one of: `cpl', `fred', or `none'. \\\\
If not `none', the following entries are included:

\TABLEFONTSIZE
\begin{itemize}
    \item \lst{lc_redchi2}: Reduced $\chi^2$ of the light curve model fit
    \item \lst{t0}: Model reference time [MJD]
    \item \lst{t1}: the time of 1\%  flux [MJD]
    \item \lst{tmax}: the time of peak flux [MJD]
    \item \lst{thalf}: the time to rise from 50\% of peak flux to peak, corrected for redshift [days]
    \item \lst{lc_rise_npts}: Number of photometric observations during the rise ($t_1 < t < t_{\rm max}$)
    \item \lst{lc_rise_ndet}: Number of photometric detections ($\geq 3\sigma$) during the rise ($t_1 < t < t_{\rm max}$)
    \item \lst{fmax_F}: peak flux in filter F [mJy]
    \item \lst{mmax_F}: \lst{fmax_F} converted to AB magnitudes [mag]
    \item \lst{Mmax_F}: \lst{mmax} converted to absolute magnitude using the distance modulus and correcting for Milky Way extinction [mag]
    \item \lst{spec_phase_min}: Minimum phase of the SNIFS spectra relative to $t_1$ or discovery, corrected for redshift [days];
    \item \lst{spec_phase_max}: Maximum phase of the SNIFS spectra relative to $t_1$ or discovery, corrected for redshift [days];
    \item \lst{spec_phase_tmax_min}: Same as \lst{spec_phase_min} except using $t_{\rm max}$ as the reference epoch [days];
    \item \lst{spec_phase_tmax_max}: Same as \lst{spec_phase_max} except using $t_{\rm max}$ as the reference epoch [days]; 
    \item \lst{phase_reference}: source of the reference time used for \lst{spec_phases}, either \texttt{lcfit} or \texttt{discovery};
\end{itemize}
\NORMALFONTSIZE

%% file: spectable_columns.tex
\TABLEFONTSIZE
\begin{itemize}
    \item \lst{dateobs}: Date of observation, in ISOT format;
    \item \lst{mjdobs}: MJD of observation;
    \item \lst{phase}: Days relative to $t_1$ or discovery;
    \item \lst{quality}: Extraction quality flag, one of: Bronze, Silver, Gold;
    \item \lst{median_snr}: Median S/N ratio of the full spectrum
    \item \lst{median_snr_blue}: Median S/N for $\lambda \leq 5100~\AAA$
    \item \lst{median_snr_red}: Median S/N for $\lambda > 5100~\AAA$
    \item \lst{exptime}: Exposure time [s]
    \item \lst{fgood}: Fraction of non-null spectrum pixels; 
    \item \lst{lbda_min}: Minimum (rest-frame) wavelength of spectrum [$\AAA$];
    \item \lst{lbda_max}: Maximum (rest-frame) wavelength of spectrum [$\AAA$];
    \item \lst{filename}: Base \textsc{ascii} filename (e.g., \lst{spec1d_18_070_017_002_lite.ascii}).
\end{itemize}
\NORMALFONTSIZE